\documentclass[journal]{IEEEtran}
\IEEEoverridecommandlockouts                              

\usepackage{color}
\usepackage{silence}
\usepackage{cite}
\usepackage{algorithmic}
\usepackage{algorithm}
\usepackage{amssymb}
\usepackage{amsmath}
\usepackage{multirow}
\usepackage{multicol}   
\usepackage[justification=centering]{caption}
\usepackage{footnote}
\makesavenoteenv{tabular}
\usepackage{graphicx}

\usepackage{caption}
\usepackage{subcaption}

\usepackage{epstopdf,float,amsfonts}
\usepackage{epsfig}

\newcommand{\qed}{\hfill \ensuremath{\square}}
\title{\LARGE \bf A Secure Learning Control Strategy via Dynamic Camouflaging for Unknown Dynamical Systems under Attacks}
\author{Sayak Mukherjee$^{1}$,
           Veronica Adetola$^{1}$ 
\thanks{$^{1}$ S. Mukherjee and V. Adetola are with the Optimization and Control Group, Pacific Northwest National Laboratory (PNNL), Richland, WA, USA. 
Emails: (sayak.mukherjee, veronica.adetola)@pnnl.gov.}}

\begin{document}

\maketitle
\begin{abstract}
      This paper presents a secure reinforcement learning (RL) based control method for unknown linear time-invariant cyber-physical systems (CPSs) that are subjected to compositional attacks such as eavesdropping and covert attack. We consider the attack scenario where the attacker learns about the dynamic model during the exploration phase of the learning conducted by the designer to learn a linear quadratic regulator (LQR), and thereafter, use such information to conduct a covert attack on the dynamic system, which we refer to as doubly learning-based control and attack (DLCA) framework. We propose a dynamic camouflaging based attack-resilient reinforcement learning (ARRL) algorithm which can learn the desired optimal controller for the dynamic system, and at the same time, can inject sufficient misinformation in the estimation of system dynamics by the attacker. The algorithm is accompanied by theoretical guarantees and extensive numerical experiments on a consensus multi-agent system and on a benchmark power grid model.  
\end{abstract}

\section{Introduction}
Security of Cyber-Physical Systems (CPSs) is becoming one of the fundamental requirements to safeguard various infrastructure and control systems against malicious attacks that can lead to catastrophic failures if left unattended. References such as \cite{pasqualetti2015control, dibaji2019systems} present overview of various theoretical and computational aspects of attack detection, prevention and resilient control designs. Extensive research work on the detection and identification of the attacks can be found in references \cite{li2016event,mousavinejad2018novel, vamvoudakis2014detection}. Various types of different attack scenarios are considered in the literature; \cite{chong2019tutorial} categorizes these scenarios based on the CPS model knowledge, disclosure of resources and disruption of resources. More specifically, recent works have investigated attacks such as denial-of-service attacks \cite{yuan2013resilient}, false data-injection attacks \cite{bai2017data}, replay attacks \cite{zhu2013performance}, covert attacks \cite{de2017covert, barboni2020detection}, etc. Different types of attacks lead to several prevention and  mitigation techniques  involving secure state estimation techniques \cite{state_est}, watermarking certain pre-specified signals in the CPS loop \cite{satchidanandan2016dynamic, mo2015physical}, moving target defense and its variants \cite{kanellopoulos2019moving, schellenberger2017detection, griffioen2020moving}, to name a few. Most of the literature on CPS security focuses on the setting where the designer knows the dynamic model of the system, and the attacker possesses the knowledge about the dynamics with varying degree of availability. However, with the ever increasing complexity and dimensionality of the dynamic systems, the designer may not have the explicit model information, and may need to learn the dynamics or the feedback controller from the state, control and output trajectories. In this paper, we propose a new formulation and mitigation technique of a compositional attack performed in a setting where the designer is tasked to learn an optimal controller in a data-driven way. We have considered both the eavesdropping and the covert attack, where the attacker can manipulate both the controls and measurements to remain undetected and at the same time harm the CPS by injecting malicious inputs, to be performed in a sequential way.
\par
Recently feedback control research for partially or fully unknown dynamic systems has seen a tremendous growth with the advancement on data-driven learning techniques such as reinforcement learning (RL) \cite{barto}. In recent years, several papers such as \cite{vrabie1,jiang1,V17,V18, MUKHERJEE_auto} have used RL for linear optimal control using a variety of solution techniques such as adaptive dynamic programming (ADP), Q-learning, actor-critic methods, model reduction based RL, etc. More variants of data-driven control research such as  data-dependent linear matrix inequalities  \cite{de2019formulas}, various distributed control designs \cite{mukherjee2020reinforcement,fattahi2020efficient}, analyzing sample complexity \cite{dean2019sample}, to name a few, have been reported. We, in this paper, consider the linear quadratic regulator problem using ADP/RL as considered in \cite{jiang1}, and then investigate the scenarios with malicious attacks. In \cite{cps_vam}, learning-based secure
control framework in the presence
of sensor and actuator attacks are discussed. This work first uses the model to perform the detection task. \cite{rangi2020learning} considers learning based attacks. We have considered a very generic attack scenario where initially the attacker does not possess any system dynamic information and therefore the attacker first eavesdrops, and after gathering sufficient dynamic information conducts a covert attack. As we consider the problem of learning the optimal control in a secured way from the perspective of the designer, we refer to our framework as a doubly learning-based control and attack (DLCA) scenario. We propose an attack-resilient reinforcement learning (ARRL) algorithm which can learn the desired optimal controller for the dynamic system, and at the same time, can inject sufficient misinformation in the system's dynamics estimation to delude  the attacker. 
\par
The contributions of this paper are as follows:
\begin{itemize}
    \item We propose a secure learning control framework, namely DLCA, where the attacker tries to exploit the learning methodology such as the system exploration to gather important dynamic information and conduct malicious covert attacks. Therefore, the vulnerability of the learning based designs for CPS in the presence of attackers needs to be addressed.
    \item Thereafter, we propose a retrofitted reinforcement learning design, namely, ARRL, where the designer tries to misguide the attacker during the exploration phase of the learning. We dynamically couple the CPS with a nonlinear time-variant static map satisfying input-to-state stability (ISS) conditions, which we termed as \textit{dynamic camouflaging}.  
    \item The proposed method is accompanied with sufficient guarantees and numerical experiments conducted on a consensus multi-agent system and on a benchmark power grid model. 
\end{itemize}
The rest of the paper is organized as follows. The secure learning control problem is introduced in Section II. The nominal ADP/RL based control design is discussed in Section III. The attack model is discussed in Section IV. Section V proposes the attack resilient design and its advantages. Numerical experiments are performed in Section VI, and concluding remarks are provided in Section VII. 
\section{Secure Learning Control Problem}
We first formalize the problem of performing secure learning control design for the CPS. We consider a linear time invariant dynamic systems with the dynamics:
\begin{align}\label{sys}
    \dot{x} = Ax + Bu, x(0)=x_0,
\end{align}
where the $x \in \mathbb{R}^n$ denotes the states, and $u \in \mathbb{R}^m$ denotes the control inputs. The designer is interested in computing the optimal controller for this dynamic systems with unknown state matrices. Therefore, we make the following assumptions from the designer's perspective:\\
\textbf{Assumption 1:} The model matrices $A$, and $B$ are unknown. \\
\textbf{Assumption 2:} The pair ($A$, $B$) is stabilizable.\\
\textbf{Assumption 3:} The state and control measurements $x(t)$, and $u(t)$ are available to the designer. 
\par
The designer is tasked with formulating a learner to solve the following linear quadratic regulator (LQR) problem.\\
\textbf{P.} Under the assumptions $1,2$ and $3$, \textit{learn} the state feedback controller $u=-Kx$ such that the following objective is minimized in closed loop:
\begin{align}\label{obj}
    \mbox{minimize} \; J(x_0,u) = \int_{0}^{\infty} (x(t)^T Q x(t) + u(t)^T R u(t))dt,
\end{align}
where $Q \succeq 0, R \succ 0$ denote the state and control penalty weights. 
\par
The standard model based solution to this problem is found by solving the well-known algebraic Riccati equation (ARE). In recent times, the research works on adaptive dynamic programming and reinforcement learning have looked into formulating techniques to solve this problem without the knowledge of the model matrices, and only using state and input trajectories (we will briefly recapitulate such nominal learning control design in the next section). However, standard learning control designs do not consider any adversarial behavior from malicious entities. In this paper, we, on the other hand, considered a scenario when the dynamic system is under the influence of an attacker. 
\par
The attacker initially does not possess any dynamic information about the system. Therefore, the attacker tries to extract the dynamic information of the system during the learning, and then conducts \textit{covert} attacks, which will be discussed shortly. Therefore, we can formalize the activities of the attacker in two phases described as follows.\\
\textbf{Attacker Act A1 (\textit{Eavesdropping}):} Attacker conducts the first phase of the attack during the \textit{exploration} phase of the learning control design. The attacker intends to learn about the dynamic matrices $A,B$ during this phase by gathering the exploration input $u(t)$, and the resultant state measurements $x(t)$. Although, the designer is conducting the exploration of the dynamic system to learn the optimal control $K$, the designer is unaware that the attacker is also using such trajectory information to learn about the system's dynamics.\\
\textbf{Attacker Act A2 (\textit{Covert Attack}):} In the next phase, the adversary conducts a covert attack on the dynamical system. In covert attacks, the attacker inject malicious signals at the inputs, and then compensate the impact of the injected attack in the measurements to conceal the attack, thereby, making it covert to the system operator. The covert attacks are very difficult to identify, and its impact on the system can be catastrophic. The attacker can also keep on injecting malicious signals at the actuation such that the  system continues to operate inefficiently without being captured by the sensors. 
\par
Therefore, the designer now needs to propose modifications to the learning control design. The attacker should not be able to eavesdrop and accurately capture the dynamic information that could allow a successful covert attack.  At the same time, the designer should learn the optimal control solutions corresponding to the dynamical system \eqref{sys} and objective \eqref{obj}. We enumerate these considerations as follows:
\begin{itemize}
    \item \textbf{Learner consideration 1:} Learner needs to make the exploration phase of the learning control secure. This means, the measurements of $u(t)$, and $x(t)$ should not be the accurate representation of the dynamical system \eqref{sys}. 
    \item \textbf{Learner consideration 2:} The attacker should be unable to keep its attack on the system \textit{covert}. Therefore, without any external disturbance, when the dynamic system operates under steady state condition, any malicious injection  will create considerable perturbations at the state measurement channels which can be easily detected by the operator. 
    \item \textbf{Learner consideration 3:} Although, the learner camouflage the exploration trajectory measurements,  it is still required to compute the optimal control $u=-Kx$ corresponding to the original learning problem $\textbf{P}$. Therefore, we do not compromise with the learning computation accuracy in order to satisfy considerations 1 and 2, which is really important from the perspective of the implemented feedback control for the dynamic system. 
\end{itemize}
We will next recapitulate the nominal adaptive dynamic programming (ADP) based optimal control learning strategy that can solve problem \textbf{P} without any malicious attack. 
\section{Recapitulation of ADP-based Nominal Learning Control}

The problem \textbf{P} 
with the unknown model matrices can be solved using ADP/ RL based approaches. 
We use the off-policy RL based gain computation framework for this study as given in \cite{jiang1}.
Here, the system is excited with exploration signal $u_0$, and, thereafter, the state measurements $x(t)$ are gathered for a sufficient number of time samples described shortly. The control input $u_0$ should be such that the state trajectory remains sufficiently bounded. 
Considering a quadratic Lyapunov function $x^TPx, P \succ 0$, the time-derivative along the state trajectories is computed, and subsequently using the Kleinman's algorithm \cite{kleinman} the following model-independent trajectory relationship can be obtained for the interval $[t,t+T]$:

\footnotesize
\begin{align}
\label{main eqn nom}
 &\hspace{-.28 cm} x^T_{(t+T)}P_kx_{(t+T)} - x^T_t P_k x_t  
 -  2\int_{t}^{t+T}((K_kx+u_{0})^TR K_{k+1}x )d\tau  \nonumber \\ & \;\;\;\;\; = -\int_{t}^{t+T}(x^T \bar{Q}_k x)d\tau.
 \end{align}
 \normalsize
 where, $\bar{Q}_{k} = Q + K_k^TRK_k $.
 We can solve \eqref{main eqn nom} by constructing a data-driven iteration framework that uses the time sampled measurements of states and the controls. The learner gathers the data matrices $\mathcal{D} = \{ \mathcal{N}_{xx}, \mathcal{M}_{xx}, \mathcal{M}_{xu_0}\}$ where,
 
 \footnotesize
 \begin{align} 
& \hspace{-.3 cm} \mathcal{N}_{xx} = \begin{bmatrix}
x \otimes x |_{t_1}^{t_1+T},& \cdots &, x \otimes x |_{t_l}^{t_l+T} 
\end{bmatrix}^T,\\
& \hspace{-.3 cm} \mathcal{M}_{xx} = \begin{bmatrix}
\int_{t_1}^{t_1+T}(x \otimes x) d\tau ,& \cdots &, \int_{t_l}^{t_l+T} (x \otimes x) d\tau \\
\end{bmatrix} ^T,\\
& \hspace{-.3 cm} \mathcal{M}_{xu_0} = \begin{bmatrix}
\int_{t_1}^{t_1+T}(x \otimes u_0) d\tau ,& \cdots & ,\int_{t_l}^{t_l+T} (x \otimes u_0) d\tau \\
\end{bmatrix} ^T.
\end{align} 

\normalsize
Algorithm 1 summarizes the steps to compute the optimal control solutions with unknown state dynamics. 

 \begin{algorithm}[]
\caption{ Nominal Reinforcement Learning  Control }
\footnotesize
1. \textit{Data gathering:}
\textit{Measure} the state and controls ($x(t)$ and $u_0$) for interval $(t_1,t_2,\cdots,t_l),t_i-t_{i-1}=T$.
Then \textit{construct}  $\mathcal{D} = \{ \mathcal{N}_{xx}, \mathcal{M}_{xx}, \mathcal{M}_{xu_0}\}$ such that rank($\mathcal{M}_{xx} \;\; \mathcal{M}_{xu_0}) = n(n+1)/2 + nm$. \\

2. \textit{Iteratively update the control :}
Starting with a stabilizing $K_0$, \textit{update} the feedback control gain $K_k$ iteratively ($k=0,1,\cdots$) by solving the least square equation where $\mbox{vec}(.)$ denotes standard vectorization operation - 

\textbf{for $k=0,1,2,..$}\\
A. \textit{Solve} for $P_k,$ and  $K_{k+1}$:

\begin{align}\label{eq:update1}
\hspace{-.3 cm} \begin{bmatrix}
\mathcal{N}_{xx} & -2\mathcal{M}_{xx}(I_n \otimes K_k^TR)  -2\mathcal{M}_{xu_0}(I_n \otimes R)
\end{bmatrix} \times\\ \nonumber 
\begin{bmatrix}
\mbox{vec}(P_k) \\ \mbox{vec}(K_{k+1} ) 
\end{bmatrix}  =-\mathcal{M}_{xx}\mbox{vec}(\bar{Q}_{k}).
\end{align}
B. \textit{Break} the iterations when $||P_k - P_{k-1}|| < \varsigma$, $\varsigma $ is a small positive threshold.\\
\textbf{endfor}\\

3. \textit{Applying K on the system :} Finally, apply $u=-Kx $, and remove $u_0$.\\
\end{algorithm} 
\normalsize
\noindent \textbf{Remark 1:} With nominally stable system, i.e., when $A$ is Hurwitz, the iterative update in \eqref{eq:update1} does not require any initial stabilizing control. Otherwise, policy iteration based RL techniques require a stabilizing $K_0$ \cite{jiang_book}, mainly because of its inception from the Newton-Kleinman updates \cite{kleinman}.\par 
\noindent \textbf{Remark 2:} In order to get unique convergent solutions of the optimal gain, the data sample requirement is converted into the following rank condition: rank($\mathcal{M}_{xx} \;\; \mathcal{M}_{xu_{0}}) = n(n+1)/2 + nm$. This is dependent on the number of unknown variables in the least squares problem. The condition is also analogous to the \textit{persistency of excitation} condition of adaptive control literature. Practically, one can use twice of the data samples required by the rank condition for guaranteed convergence. 
\par
\noindent \textbf{Theorem 1 \cite{jiang1}:} When the rank condition of the Remark 2 is satisfied, the iterates of $P_k, K_k$ from Algorithm 1 converge to optimal $P$, and $K$ with $k \to \infty$. \qed 
\vspace{-.35 cm}
\section{Attacker Modeling}
To this end, we consider an attacker which acts in two folds as follows.
\vspace{-.45 cm}
\subsection{Attacker Act 1 - Eavesdropping:} The attacker becomes active during the exploration phase of the learning algorithm. In this phase, the attacker eavesdrops on the input and state channels $u_0(t)$, and $x(t)$ to gather the measurements for the time instants $t_1,...,t_l$. The attacker gathers as much as dynamic information as possible to perform the identification of $A, B$. The attacker can employ any of the sub-space based identification approaches. We assume that the attacker knows the dimensionality of the system. The attacker constructs the surrogate model of the original dynamical system as follows:
\begin{align}
    \dot{\tilde{x}} = \tilde{A}\tilde{x} + \tilde{B}u, \tilde{x}(0) =\tilde{x}_0.
\end{align}
It can be expected that the model matrices $\tilde{A}, \tilde{B}$ are identified with high accuracy. Practically, the attacker can identify a model which is similar to the original state space, i.e, they are related by a similarity transformation $T$:
\begin{align}
    \tilde{A} = TAT^{-1}, \tilde{B} = TB.
\end{align}
In order to consider the \textit{worst-case scenario}, we assume that the attacker can successfully identify the actual state space models. Therefore, $\tilde{A} = A, \tilde{B} = B$, i.e., $T=I$. \\
\textbf{Remark 3:} \textit{Why eavesdropping during exploration?}  The reason behind such consideration is that the exploration is the most vulnerable phase for the system to be compromised, creating a worst-case scenario. During exploration, the designer tries to persistently excite the system such that the input-output data is sufficiently rich in dynamic information. Therefore, if the attacker can get access to such exploration data, the attacker could easily perform the system identification to get accurate model estimates.
\vspace{-.4 cm}
\subsection{Attacker Act 2 - Covert Attack:} Once, the attacker has access to the system dynamic model information, a \textit{covert attack} using the manipulated input and state channels is conducted. An attack on the control input will result into:
\begin{align}\label{att_input}
    \tilde{u}(t) = u(t) + \zeta (t),
\end{align}
where $\zeta (t)$ is the malicious input signal. 
Thereafter, the attacker compensates the effect of the input attack to the state measurement sensors of the system by manipulating:
\begin{align}\label{att_meas}
    \bar{x}(t) = x(t) - \tilde{x}(t),
\end{align}
where $\bar{x}(t)$ is the measured states, and $\tilde{x}(t)$ is the compensation signals added by the attacker. 
Therefore, the attacker needs to generate $\zeta(t)$ and $\tilde{x}(t)$. We now state the following Lemma which characterizes the covert nature of the attack.
\noindent \textbf{Lemma 1:} 
Consider the attack begins at $t = T_a$, if the attacker uses \eqref{att_input}, \eqref{att_meas}, and runs the dynamics:
\begin{align}\label{attack_model}
    \mathcal{A} : \;\; &\dot{\tilde{x}} = A\tilde{x} + B\zeta,
\end{align}
with $\tilde{x}(T_a) = 0$, then the attack will remain covert at the measured states.\\ 
\\
Proof: The solution of $x(t)$ starting from $t=T_a$ is given as:
\begin{align}
    x(t)&=  e^{A(t-T_a)}x(T_a) + \int_{T_a}^t e^{A(t - \tau)} [B(u + \zeta) ] d\tau,
\end{align}
and the output of the attacker's internal model \eqref{attack_model} is given by,
\begin{align}
    \tilde{x}(t) =  e^{A(t-T_a)}\tilde{x}(T_a) + \int_{T_a}^t e^{A(t - \tau)} [B\zeta ] d\tau.
\end{align}
Therefore, the modified output seen by the designer following from \eqref{att_meas}:
\begin{align}
    \bar{x}(t) =  e^{A(t-T_a)}(x(T_a) - \tilde{x}(T_a)) + \int_{T_a}^t e^{A(t - \tau)}[Bu] d\tau
\end{align}
Setting $\tilde{x}(T_a) =0$, 
it is evident that $\bar{x}(t)$ can be treated as legitimate state measurements, making the attack covert. 
\qed

\section{Retrofitting the Learning to make it Attack-Resilient}
\begin{figure*}[]
    \centering
    \includegraphics[width=.9\textwidth, trim=4 4 4 4]{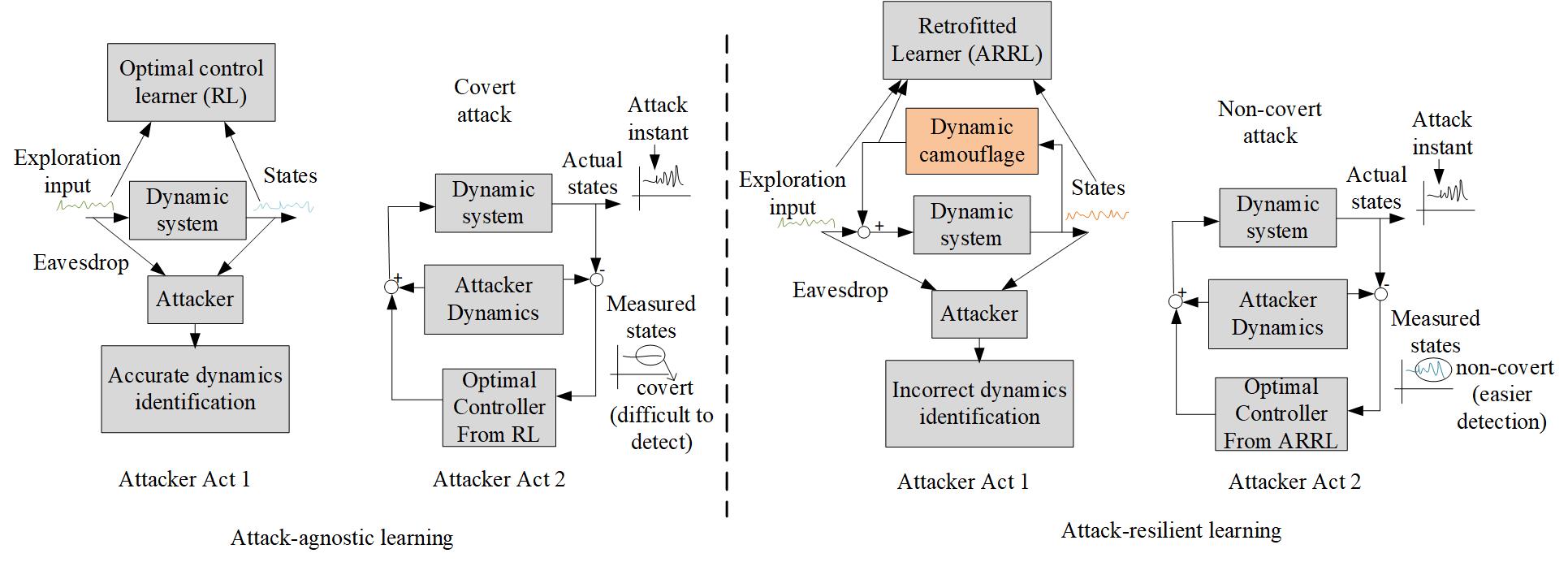}
    \caption{Overview of the attack resilient learning methodology}
    \label{fig:my_label}
\end{figure*}
The designer has to make sure that the system identification by the attacker is incorrect, and at the same time, the desired feedback control gain is being computed without sacrificing performance as enumerated in Section II. To this end, we, hereby, propose a solution in this section. The main idea is to modify the dynamic model in such a way that the attacker cannot perfectly identify the system during the exploration phase of the learning, referring to as \textit{dynamic camouflaging}. 
On the other hand, the designer has full knowledge about this extraneous modification such that the optimal gain $K$ can be correctly learned as in \textbf{P}.  
\par
As the designer starts the learning with the assumption of unknown $A,$ and $B$ matrices, any system parameter or actuator gains cannot be modified. Instead, we suggest to add an input-to-state stable (ISS) coupling, dependent on the states of the dynamic system, at an internal actuation location of the system, which is assumed to be safe with respect to external attacks. To check the applicability of such coupling, operators can use the plant simulators, or historic measurements to estimate whether the state trajectories will remain bounded. The underlying dynamic system, therefore, is modified to:
\begin{align}\label{modified}
    &\dot{x} = Ax + B(u + \psi),\\
    &\psi = \phi(t,x(t)).
\end{align}
Here $\psi$ is assumed to be a nonlinear time varying static map, however, one can also make this a dynamic map. We make the following boundedness assumption on the coupling function $\psi = \phi(t,x(t))$.\\
\noindent \textbf{Assumption 4:} The coupling function $\psi = \phi(t,x(t))$ satisfies
\begin{align}
    \| \psi \|_2 = \| \phi(t,x(t)) \|_2 \leq \gamma \| x(t) \|_2, \gamma >0.
\end{align}
Thereafter, we characterize the ISS stability of the interconnection during the exploration such that the state trajectories remain bounded.\\

\noindent \textbf{Lemma 2:} With bounded exploration inputs, and under assumption 4, the system \eqref{modified} will be input-to-state stable (ISS) with respect to $\psi$.

\noindent \textbf{Proof:} For any generic dynamic system the exploration control can be written in the form $u=-K_{s}s + u_0$ where $(A-BK_s)$ is stable, and $u_0$ is a bounded exploration such that the state measurements remain within the stable neighborhood of the operating point. Therefore, we could write,  
\begin{align}
    \dot{x} = (A-BK_s)x + Bu_0 + B\psi(t), x(t_0)=x_0.
\end{align}
The resulting state trajectories become 
\begin{align}
    x(t) = e^{(A-BK_s)(t-t_0)}x_0 + \int_{t_0}^{t}e^{(A-BK)(t-\tau)}Bu_0 d\tau \\ \nonumber + \int_{t_0}^{t}e^{(A-BK)(t-\tau)}B\psi(\tau) d\tau
\end{align}
As $A-BK_s$ is Hurwitz, we have, $\| e^{(A-BK_s)(t-t_0)}\| \leq ke^{-\lambda (t-t_0)}, k>0,\lambda > 0$. Thereafter, we can bound $\| x(t)\|$ as,
\begin{align}
    & \| x(t)\| \leq ke^{-\lambda (t-t_0)}\| x(t_0)\| + \\ \nonumber & \frac{k\|B\|}{\lambda} (\mbox{sup}_{\tau \in [t_0,t]} \|u_0(\tau)\| + \mbox{sup}_{\tau \in [t_0,t]} \|\psi(\tau)\|).
\end{align}
Therefore, we can conclude that with bounded $\| u_0(t) \|$ , and assumption $4$, the $x(t)$ dynamics remains ISS. \qed 
\par
This type of dynamics has been recently studied in our papers \cite{mukherjee2020robust, mukherjee2020imposing} in the context of robustness of structured control designs. Motivated from them, the formalism is found to be suitable to make the nominal RL algorithm attack-resilient. When the attacker eavesdrops on the input and state channels, then $u(t)$ and $x(t)$ correspond to the underlying dynamics \eqref{modified}, and, therefore, performing a system identification will result into erroneous state space representations, i.e.,
\begin{align}
    \tilde{A} \neq TAT^{-1}, \tilde{B} \neq TB.
\end{align}
In our numerical experiments we set $\phi(t,x(t)) = f(t)x(t)$ with $\| f(t) \|_2 \leq \gamma$, $f(t) \neq 0, \forall t$, and to simulate the worst-case, we assume that the identified state matrix is $\tilde{A}= A + Bf(t)$, with $f(t)$ frozen at time $t=t_0$. We assume that the attacker could perfectly estimate $B$, to create a worst-case scenario.   

To this end, the learning algorithm needs to be retrofitted due the modifications as in \eqref{modified}. As we have intentionally connected the coupling $\psi$, the designer has access to the full measurements of $\psi(t)$. At this stage recall the Kleinman's algorithm:

\noindent \textbf{Theorem 2 \cite{kleinman}: } \textit{Let $K_0$ be such that $A-BK_0$ is Hurwitz. Then, for $ k=0,1,\dots $ \\
1. Solve for ${P}_k$(Policy Evaluation) :
\begin{align}\label{Kleinman1}
\hspace{-.3 cm} &A_{k}^T{P}_k + {P}_k A_{k} + {K}_k^TR{K}_k + Q = 0, A_{k} = A-B{K}_k.
\end{align}
2. Update the control gain (Policy update):
\begin{align}\label{Kleinman2}
{K}_{k+1} = R^{-1}B^T {P}_k.
\end{align}
Then $A - BK$ is Hurwitz and $K_{k} $ and $P_{k}$ converge to optimal $K$, and $P$ as $ k  \rightarrow \infty $.} \qed \\
The modified state dynamics \eqref{modified} incorporating $u = -{K}_kx$ is given by
\begin{align}
\dot{x} &= (A - B{K}_k)x + B({K}_kx + u) + B\psi.
\end{align} 

We use similar exploration inputs $u=u_0$ as before. As considered in the derivation of Algorithm 1, we consider similar Lyapunov function candidate  $x^T{P}_kx$, and compute its derivative along the closed-loop trajectories and use Theorem 2 to replace the dependency on the model matrices as follows:
\begin{align}
&\frac{d}{dt}(x^T{P}_kx) = x^T(A_{k}^T{P}_k + {P}_kA_{k})x + \nonumber \\ 
& \;\;\;\;\; 2({K}_kx+u_{0})^TB^T{P}_kx + 2(\psi^TB^T{P}_k)x \\ 
&  = -x^T({Q}_{k} )x + 2({K}_kx+u_{0})^TR{K}_{k+1}x + \nonumber \\  & \hspace{3 cm} 2(\psi^TB^T{P}_k)x, \nonumber 
\end{align}
where, ${Q}_{k} = Q + {K}_k^TR{K}_k$. Rearranging, and taking integrals on the both sides, we have,
\begin{align}
\label{main eqn}
 &\hspace{-.3 cm} x^T(t+T){P}_kx(t+T) - x^T(t) {P}_k x(t) \nonumber \\
 & -  2\int_{t}^{t+T}(({K}_kx+u_{0})^TR {K}_{k+1}x )d\tau  \nonumber \\ 
 & = -\int_{t}^{t+T}(x^T {Q}_k x - 2(\psi^TB^T{P}_k)x ) d\tau.
\end{align} 
\normalsize
\eqref{main eqn} is independent of model matrices, and constructed by the trajectories of system states $x(t)$, exploration control $u_{0}(t)$, and the measurements of the coupling $\psi(t)$. We can use the properties of Kronecker product (denoted by $\otimes$) to write $x^T{Q}_kx = (x^T \otimes x^T)\,\mbox{\mbox{vec}}({Q}_k), \psi^TB^T{P}_kx = (x^T \otimes \psi^T)\,\mbox{\mbox{vec}}(B^T{P}_k)$.
The Attack-resilient RL algorithm can be written by formulating an iterative version of \eqref{main eqn}, using measurements of $x(t),u_{0}(t)$ and $\psi(t)$ as given in Alg. 2. 
We append the data matrix $\mathcal{D}$ as in Algorithm 1 with $\mathcal{M}_{x\psi}$  where,
\begin{align} 
& \hspace{-.3 cm} \mathcal{M}_{x\psi} = \begin{bmatrix}
\int_{t_1}^{t_1+T}(x \otimes \psi) d\tau ,& \cdots & ,\int_{t_l}^{t_l+T} (x \otimes \psi) d\tau \\
\end{bmatrix} ^T.
\end{align}
\normalsize

\begin{algorithm}[]
\footnotesize
\caption{ Attack-Resilient RL (ARRL) Control }
1. \textit{Data gathering:}
\textit{Measure} the state and controls and coupling variables ($x(t)$ $u_0(t)$ and $\psi(t)$) for interval $(t_1,t_2,\cdots,t_l),t_i-t_{i-1}=T$.
Then \textit{construct}  $\mathcal{D} = \{ \mathcal{N}_{xx}, \mathcal{M}_{xx}, \mathcal{M}_{xu_0}, \mathcal{M}_{x\psi}\}$ such that rank($\mathcal{M}_{xx} \;\; \mathcal{M}_{xu_0} \;\; \mathcal{M}_{x\psi})= n(n+1)/2 + 2nm$. \\

2. \textit{Iteratively update the control :}
Starting with a stabilizing $K_0$, \textit{update} the feedback control gain $K$ iteratively ($k=0,1,\cdots$) by solving the least square equation - 

\textbf{for $k=0,1,2,..$}\\
A. \textit{Solve} for $P_k,$ and  $K_{k+1}$:

\begin{align}\label{eq:update} \underbrace{\begin{bmatrix}
\mathcal{N}_{xx} & -2\mathcal{M}_{xx}(I_D \otimes {K}_k^TR)  -2\mathcal{M}_{xu_{i0}}(I_D \otimes R) & -2\mathcal{M}_{x\psi}
\end{bmatrix}}_{\Theta_{k}} \\ \nonumber \times \begin{bmatrix}
\mbox{vec}({P}_{k}) \\ \mbox{vec}({K}_{(k+1)}) \\ \mbox{vec}(B^T{P}_{k})
\end{bmatrix} =\underbrace{-\mathcal{M}_{xx}\mbox{vec}(Q_{ik})}_{\Phi_{k}}.
\end{align}
B. \textit{Break} the iteartions when $||P_k - P_{k-1}|| < \varsigma$, $\varsigma $ is a small positive threshold.\\
\textbf{endfor}\\

3. \textit{Applying K on the system :} Finally, apply $u=-Kx $, and remove $u_0$.\\
\end{algorithm} 
\normalsize

\noindent \textbf{Theorem 3:} Performing Algorithm 2 with the modified system \eqref{modified} will able to recover the optimal control $u=-Kx$ for the actual system (1).\\
\noindent \textbf{Proof:} Performing Algorithm 2 (ARRL) using $x(t), u(t)$, and $\psi(t)$ is equivalently solving the trajectory relationship \eqref{main eqn}. As \eqref{main eqn} has been constructed using Theorem 2, then any solution from Theorem 2 will satisfy the $k^{th}$ iteration of the following equation:
\begin{align}\label{lemma2}
    \Theta_{k}\begin{bmatrix}
\mbox{vec}({P}_{k}) \\ \mbox{vec}({K}_{(k+1)}) \\ \mbox{vec}(B^T{P}_{k})
\end{bmatrix} =\Phi_{k}.
\end{align}
Therefore, a solution from Theorem 2 should also satisfy \eqref{lemma2}. With sufficient gathering of data, the condition rank($\mathcal{M}_{xx} \;\; \mathcal{M}_{xu_{0}} \;\; \mathcal{M}_{x \psi}) = n(n+1)/2 + 2nm$ is satisfied, and, therefore, $\Theta_{k}$ will have full column rank. As such, equation \eqref{lemma2} has a unique solution ${P}_{k}, {K}_{(k+1)}.$ As this is an unique solution, it is also the solution ${P}_{k}, {K}_{k+1}$ of theorem 2. Considering this equivalence of the Algorithm 2 with the modified Kleinman update in Theorem 2, we can conclude that the ${K}$, and ${P}$ corresponding to Algorithm 1 can be recovered. \qed

The attacker then tries to launch covert attacks once the closed-loop system is in the operating condition. 
However, the measured states seen by the designer now is:
$\bar{x}(t) = x(t) - \tilde{x}(t)$:
\begin{align}
    &\bar{x}(t) =  e^{A(t-T_a)}x(T_a) - e^{\tilde{A}(t-T_a)}\tilde{x}(T_a) + \\ \nonumber 
    &\int_{T_a}^t ( e^{A(t - \tau)} [B(u + \zeta)] - e^{\tilde{A}(t - \tau)} [\tilde{B}\zeta ] ) d\tau,
\end{align}
which is not a legitimate system response, even if $\tilde{B}=B$, and therefore will create undesired perturbations during the normal operational mode of the plant depending on the energy in attack inputs, and the error $\| A- \tilde{A}\|$. Therefore, a pre-tuned set-point based detector can alert the system operator to take further necessary actions. The designer can also design sophisticated detectors using the nominal statistical properties of the system, which we keep as future research.
\vspace{-.4 cm}
\section{Numerical Simulations}
\subsection{A Multi-agent System}
\begin{figure*}
    \centering
    \begin{minipage}{0.3\linewidth}
    \centering
    \includegraphics[width = \linewidth, height =2.7 cm]{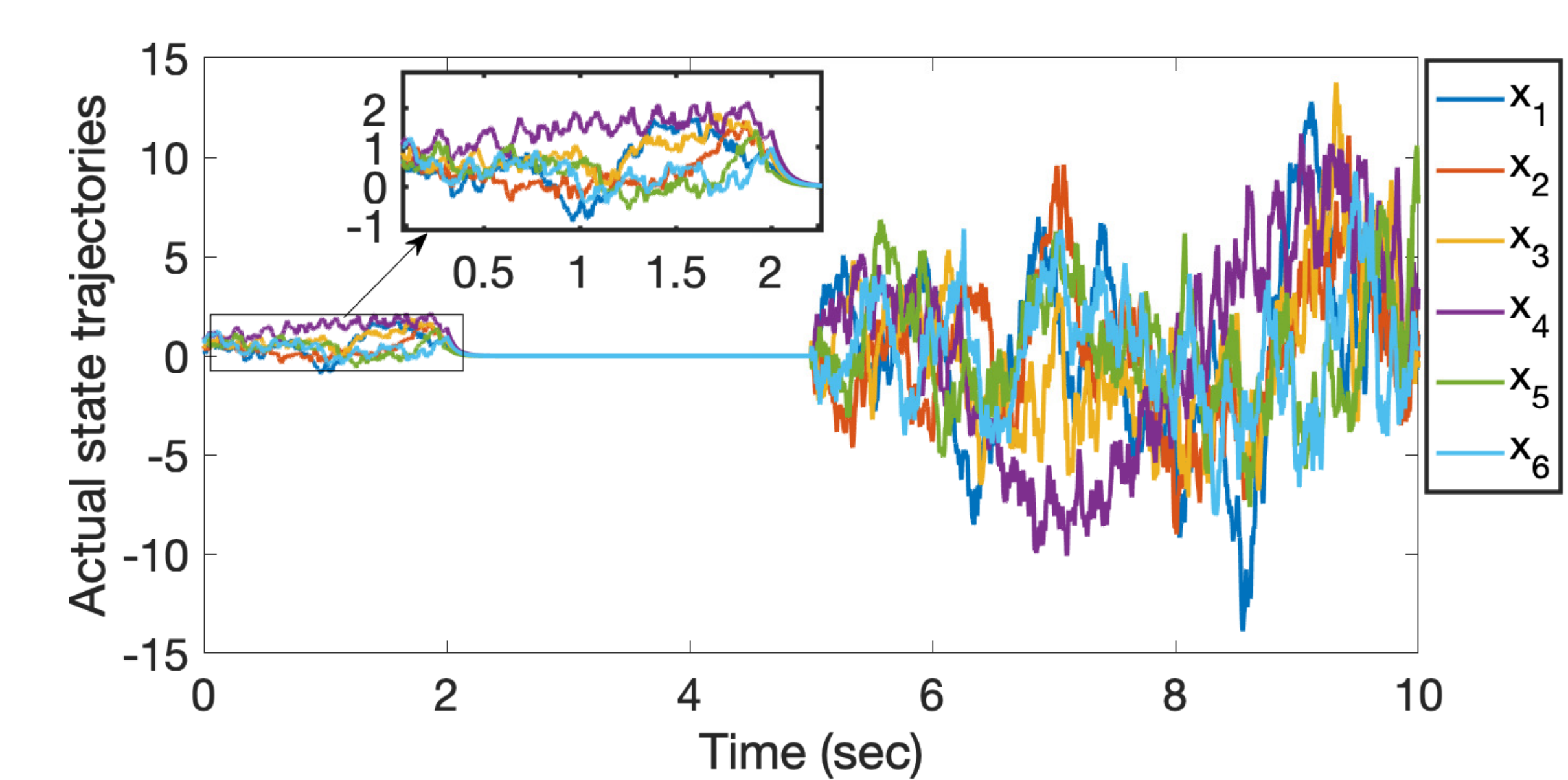}
    \caption{\small{Actual states during exploration (till 2 s), control implementation and attack at 5 s for the nominal design}}
    \label{fig:act_covert}
    \end{minipage}
    \begin{minipage}{0.3\linewidth}
    \includegraphics[width = \linewidth, height =2.7 cm]{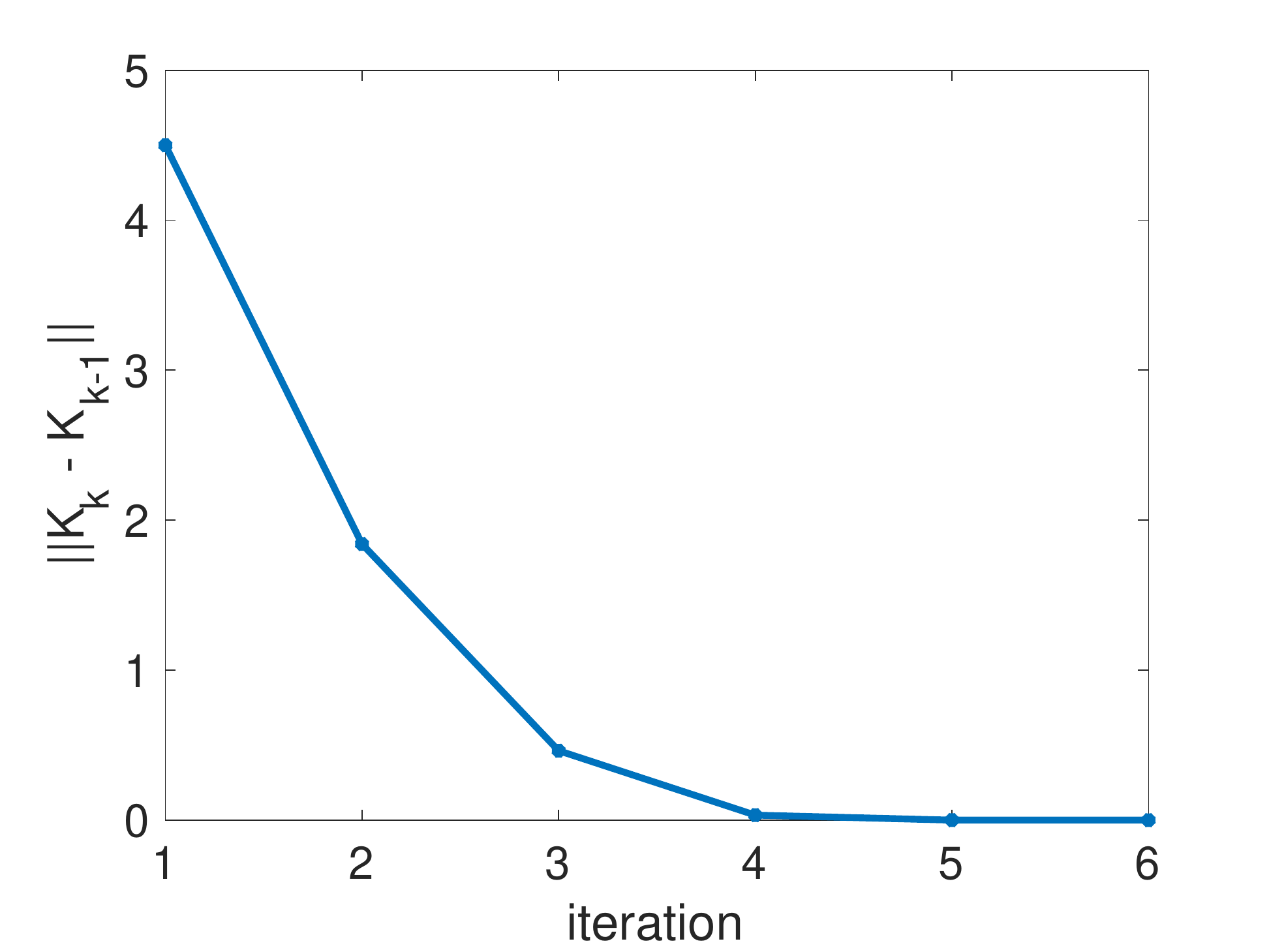}
     \caption{ \small{Convergence of controller $K$ in Algorithm 1 for the nominal design\\}}
        \label{fig:covert_K}
    \end{minipage}
    \begin{minipage}{0.3\linewidth}
        \centering
        \includegraphics[width = \linewidth, height =2.7 cm]{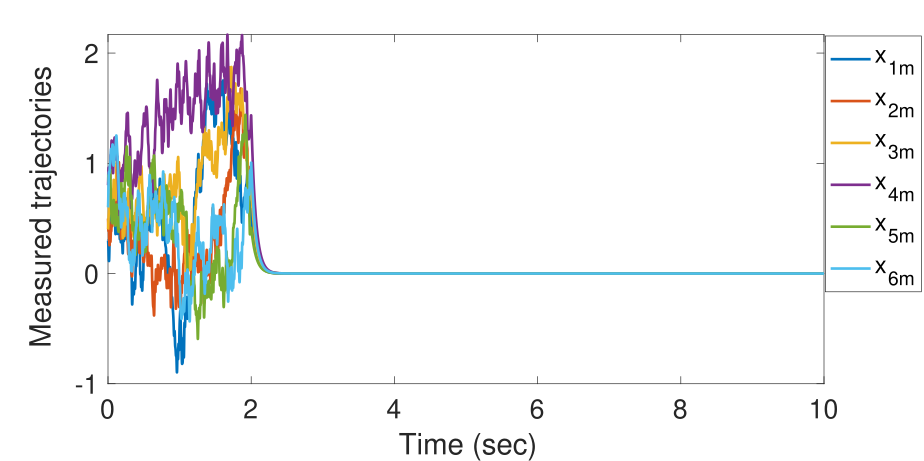} 
        \caption{\small{Measured state trajectories during exploration, control implementation, and attack at 15 s without resilient design}}
        \label{fig:meas_covert}
    \end{minipage}
    \begin{minipage}{0.3\linewidth}
        \centering
        \includegraphics[width = \linewidth, height =2.7 cm]{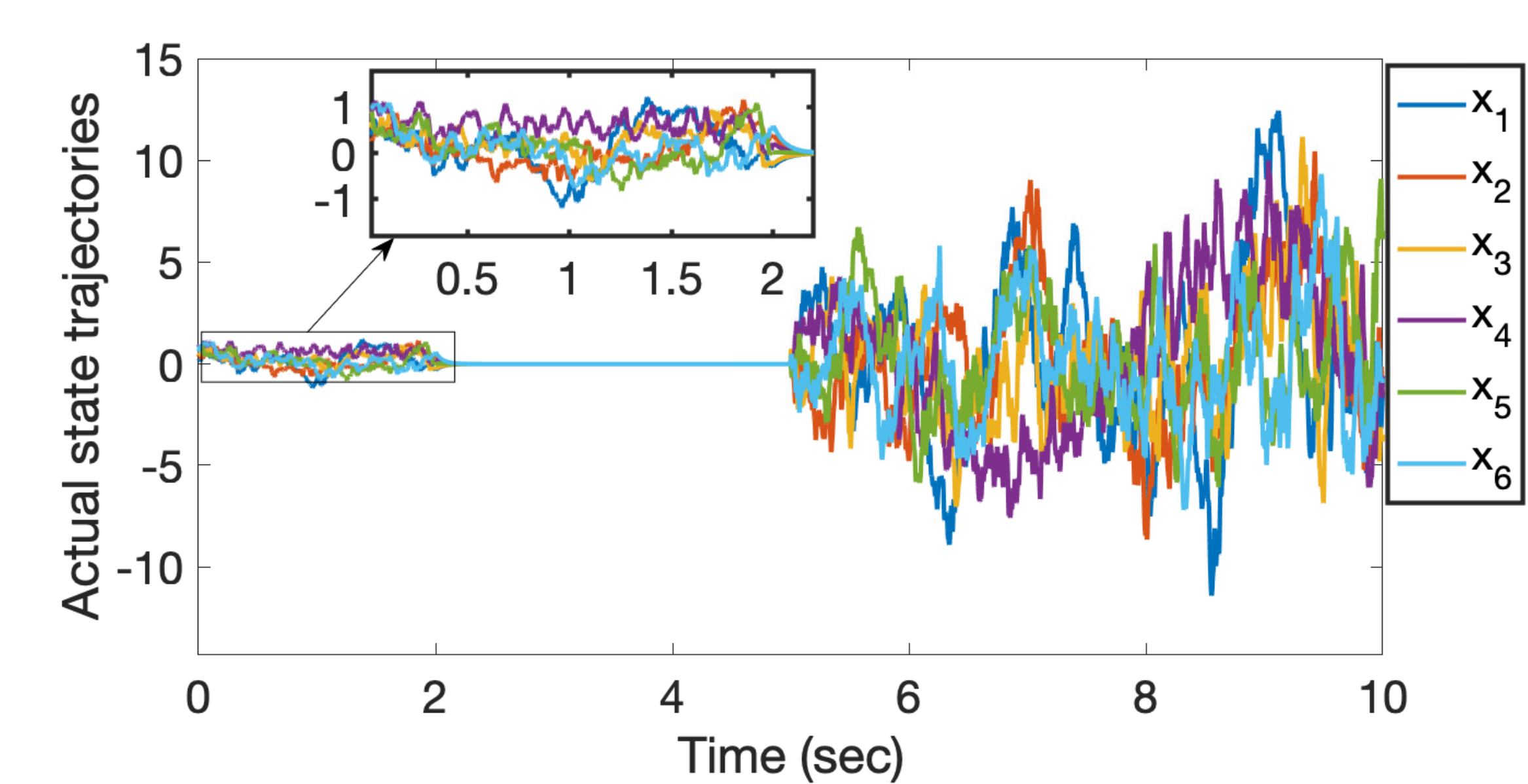} 
        \caption{\small{Actual states during exploration (till 2 s), control implementation and attack at 5 s for the resilient design\\}}
        \label{fig:attack_signal_covert}
    \end{minipage}
    \begin{minipage}{0.3\linewidth}
        \centering
        \includegraphics[width = \linewidth, height =2.7 cm]{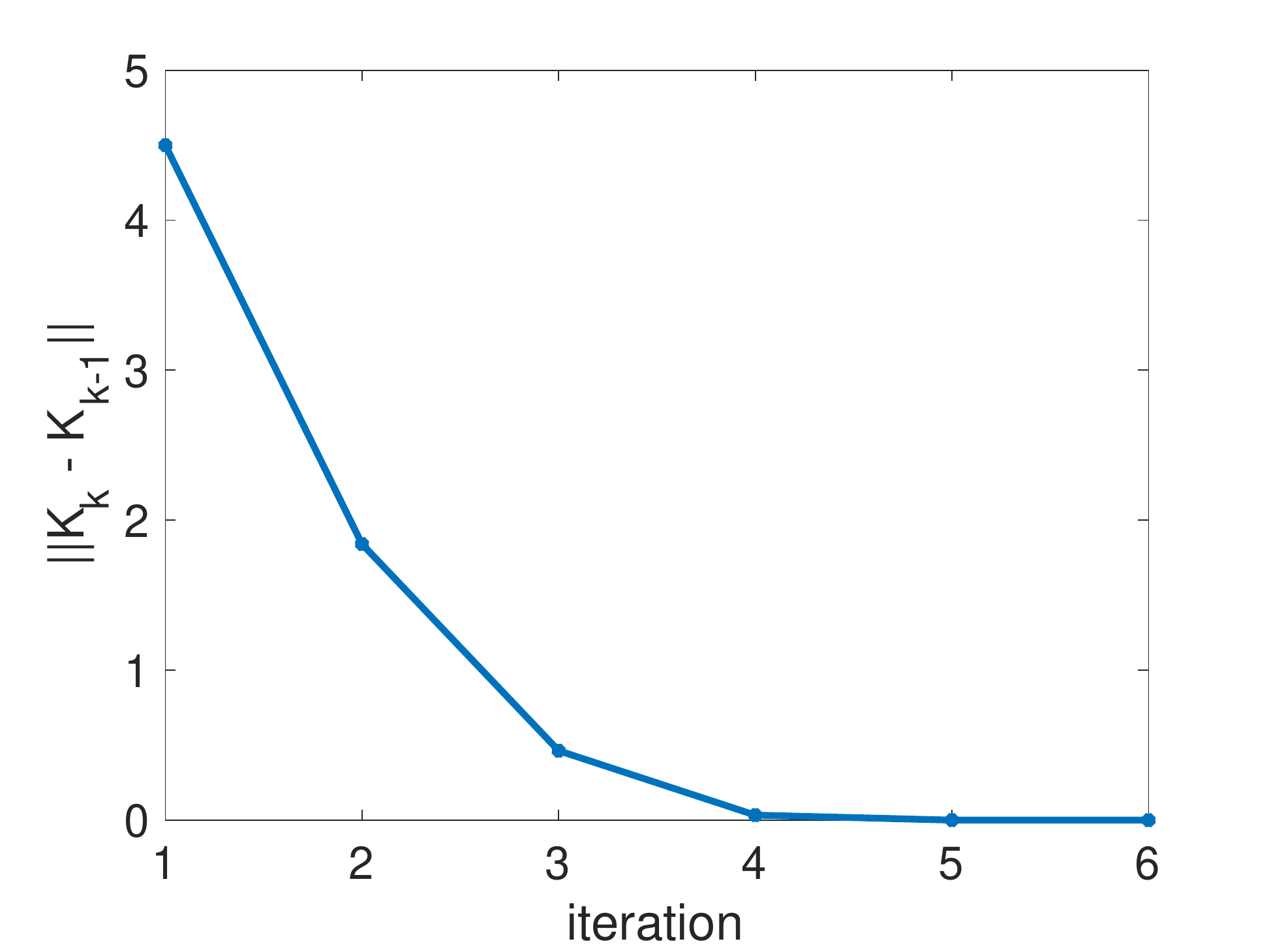} 
        \caption{\small{Convergence of the retrofitted controller following ARRL in Algorithm 2}}
        \label{fig:noncovert_K}
    \end{minipage}
    \begin{minipage}{0.3\linewidth}
        \centering
        \includegraphics[width = \linewidth, height =2.7 cm]{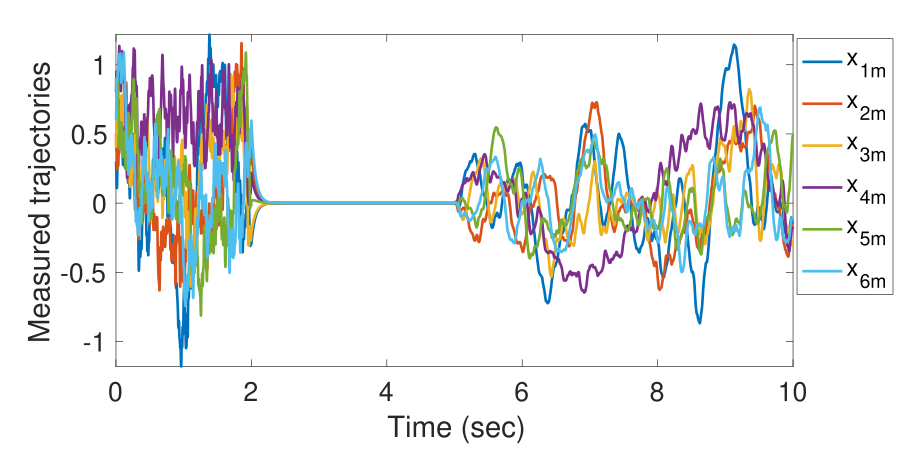} 
        \caption{\small{Measured state trajectories during exploration, control implementation, and attack at 5 s with ARRL}}
        \label{fig:meas_noncovert}
    \end{minipage}
   \vspace{-.5 cm}
\end{figure*}
We consider a multi-agent network with $6$ agents taken from \cite{mukherjee2020imposing} where each agent follows the consensus dynamics:
\begin{align}
    \dot{x}_i = \sum_{j \in \mathcal{N}_i, i \neq j} \alpha_{ij}(x_j - x_i) + u_i, x_i(0) = x_{i0}, 
\end{align}
where $\alpha_{ij} > 0$ are the coupling coefficients. We consider the state and input matrix to be:
\begin{align}
\footnotesize
    A = \begin{bmatrix} -5 &2& 3& 0& 0& 0\\
     2 &-6 &0& 0& 1& 3\\
     3 &0& -5& 2 &0 &0\\
     0 &0& 2& -2& 0& 0\\
     0 &1& 0 &0 &-4 &3\\
     0 &3& 0 &0& 3& -6 \end{bmatrix}, B=I_6.
\end{align}
\normalsize
The dynamics follows a Laplacain structure with $A.\mathbf{1_n} = \mathbf{0}$ resulting into a zero eigenvalue with the rest of them are $-10.00,
   -8.27,
   -6.00,
   -3.00,
   -0.72,
   -0.00$. We choose initial conditions as $[0.3,0.5,0.4,0.8,0.9,0.6]^T$. The learned controller is tasked to improve the damping of the slow eigenvalues. We set $Q = 10I_6, R=I_6$.
 We first experiment with the nominal system without any retrofitting in the learning design. As, $n=6, m=6,$ the rank condition for the algorithm 1 requires $2\times(57) = 114$ samples. The exploration has been performed with $200$ samples with $0.01$ s time step. We excite the system with the sum of sinusoids exploration. We assume that this phase is not secured to attackers, and therefore, the attacker could eavesdrop and could easily estimate the state matrices. Once the data matrix $\mathcal{D}$ is constructed, the control gains are computed via the iterations as given in Algorithm 1, and then implemented to the system. Fig. \ref{fig:act_covert} shows the actual state trajectories during the initial $2$ s exploration, and with the control implementation. Fig. \ref{fig:covert_K} shows the convergence of the learning control using Algorithm $1$ resulting into:
 
 \scriptsize
 \begin{align}
     K_{Alg. 1} = \begin{bmatrix}
     2.3868  &  0.2731 &   0.4239 &   0.0342    &0.0125&    0.0318\\
    0.2731  &  2.2564    &0.0319&    0.0010   & 0.1899 &   0.4100\\
    0.4239   & 0.0319   & 2.3884 &   0.3161  &  0.0004  &  0.0017\\
    0.0342   & 0.0010  &  0.3161  &  2.8112 &  -0.0001  & -0.0001\\
    0.0125    &0.1899 &   0.0004   &-0.0001 &   2.5188  &  0.4408\\
    0.0318    &0.4100&    0.0017   &-0.0001&    0.4408  &  2.2781
     \end{bmatrix},
 \end{align}
 \normalsize
 which matches closely with the model-based solution. At $5$ s, the attacker starts injecting malicious signals to the system, however, as the learning control was not secured, the attacker could launch a covert attack, and therefore, the state measurements could not able to capture any of these malicious signals as shown in Fig. \ref{fig:meas_covert}. However, the actual state trajectories are heavily impacted as shown in Fig. \ref{fig:act_covert}. This kind of covert attack can cause expensive state excursions, and make the system less efficient. The quadratic cost incurred from 5 s to 10 s turns out to be $1.08\times 10^3$ units. 

Thereafter, we show the efficacy of the retrofitted secured learning design. During the exploration phase of the learning, we have added a functional coupling with the control inputs in the form, $u = u_{control} - f(t)x(t)$, $f(t) = 0.3\times \mbox{(sin(t) + cos(t) + 0.02)}$. As the attacker does not aware of such modifications, the system identification performed by the attacker using the input and state measurements during the exploration will be erroneous. The modified algorithm 2, however, is tasked with computing the optimal control $u = -Kx$ associated with the actual system dynamics $A,B$, and not that of the modified state dynamics. The control computed with Algorithm 2  with the convergence shown in Fig. \ref{fig:noncovert_K} is given as:

\scriptsize
\begin{align}
    K_{Alg. 2} = \begin{bmatrix}
    2.3868 &   0.2731   & 0.4239&    0.0342   & 0.0125 &   0.0318 \\
    0.2731 &   2.2564   & 0.0319 &   0.0010   & 0.1899  &  0.4100\\
    0.4239  &  0.0319   & 2.3884  &  0.3161   & 0.0004  &  0.0017\\
    0.0342   & 0.0010  &  0.3161   & 2.8112  & -0.0001 &  -0.0001\\
    0.0125   & 0.1899 &   0.0004 &  -0.0001 &   2.5188  &  0.4408\\
    0.0318   & 0.4100&    0.0017  & -0.0001&    0.4408   & 2.2781
    \end{bmatrix},
\end{align}
\normalsize
which shows that the modified algorithm does not suffer in the accuracy of computing the desired optimal control. As the attacker is not able to capture the accurate state dynamics, the attack does not remain covert anymore. To simulate a worst-case identification scenario, assume that the model identified by the attacker uses $\tilde{A} = A - \epsilon_{sc} \times 0.3\times 1.02 \times I_6$ with $f(t=0)$, and $\epsilon_{sc} = 2$ is a scaling factor. Thereafter, when the attacker injects malicious attack signal at $t=5 s$, the state trajectories at the measurement ports can capture such behaviours as shown in Fig. \ref{fig:meas_noncovert}.  If the state trajectories hit a pre-calibrated set point, then the system operator is being alerted to remove the malicious enterprise.
\vspace{-.4 cm}
\subsection{A Power Grid Benchmark}
\begin{figure}[]
    \centering
    \includegraphics[width = .8\linewidth, height =4 cm]{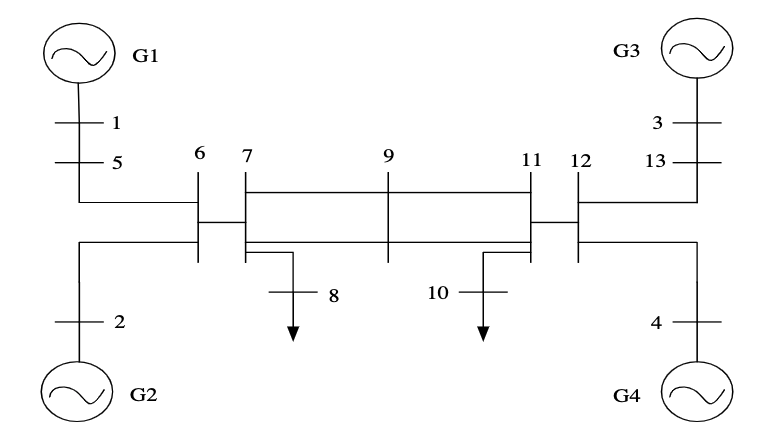}
    \caption{\small{Kundur $13$-bus $4$-machine power system model}}
    \label{fig:Kundur}
    \vspace{-.58 cm}
  \end{figure}
\begin{figure*}[]
    \begin{minipage}{0.3\linewidth}
    \includegraphics[width = \linewidth, height =2.7 cm]{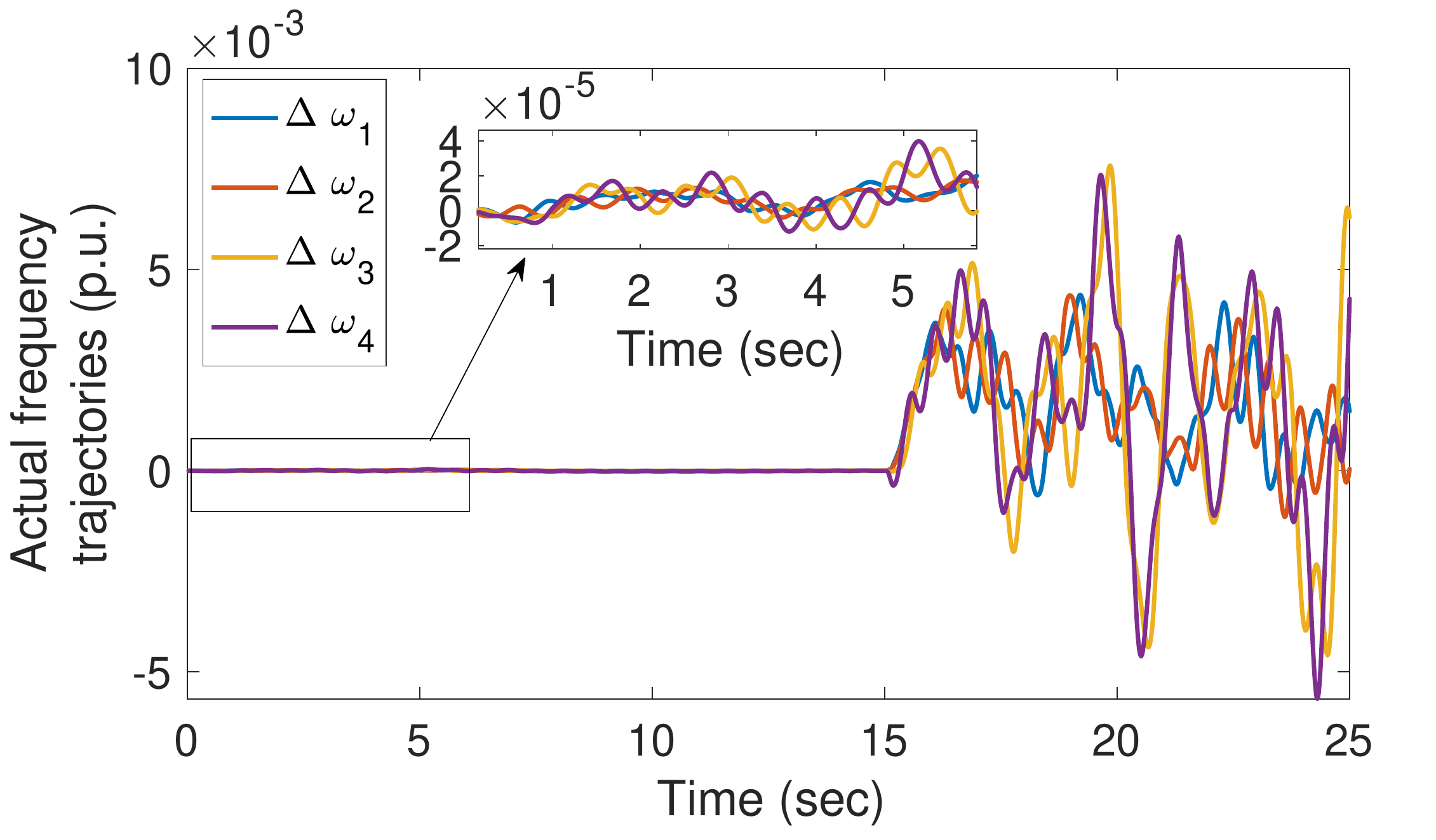}
     \caption{ \small{Actual frequencies during probing (till 5 s), control implementation and attack at 15 s without ARRL}}
        \label{fig:grid_covert_act}
    \end{minipage}
    \begin{minipage}{0.3\linewidth}
        \centering
        \includegraphics[width = \linewidth, height =2.7 cm]{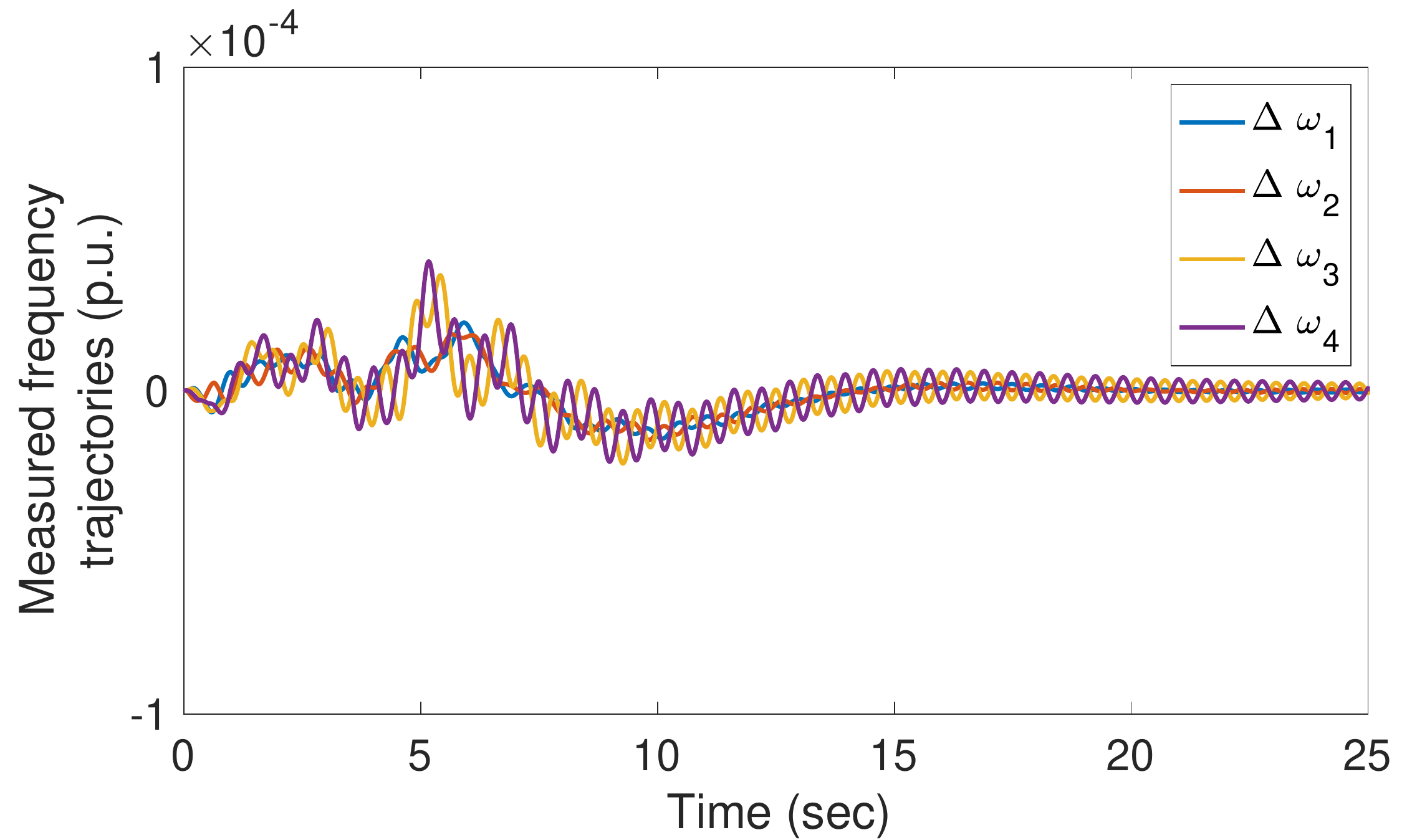} 
        \caption{\small{Measured frequencies during probing (till 5 s), control implementation and attack at 15 s without ARRL}}
        \label{fig:grid_covert_meas}
    \end{minipage}
    \begin{minipage}{0.3\linewidth}
        \centering
        \includegraphics[width = \linewidth, height =2.7 cm]{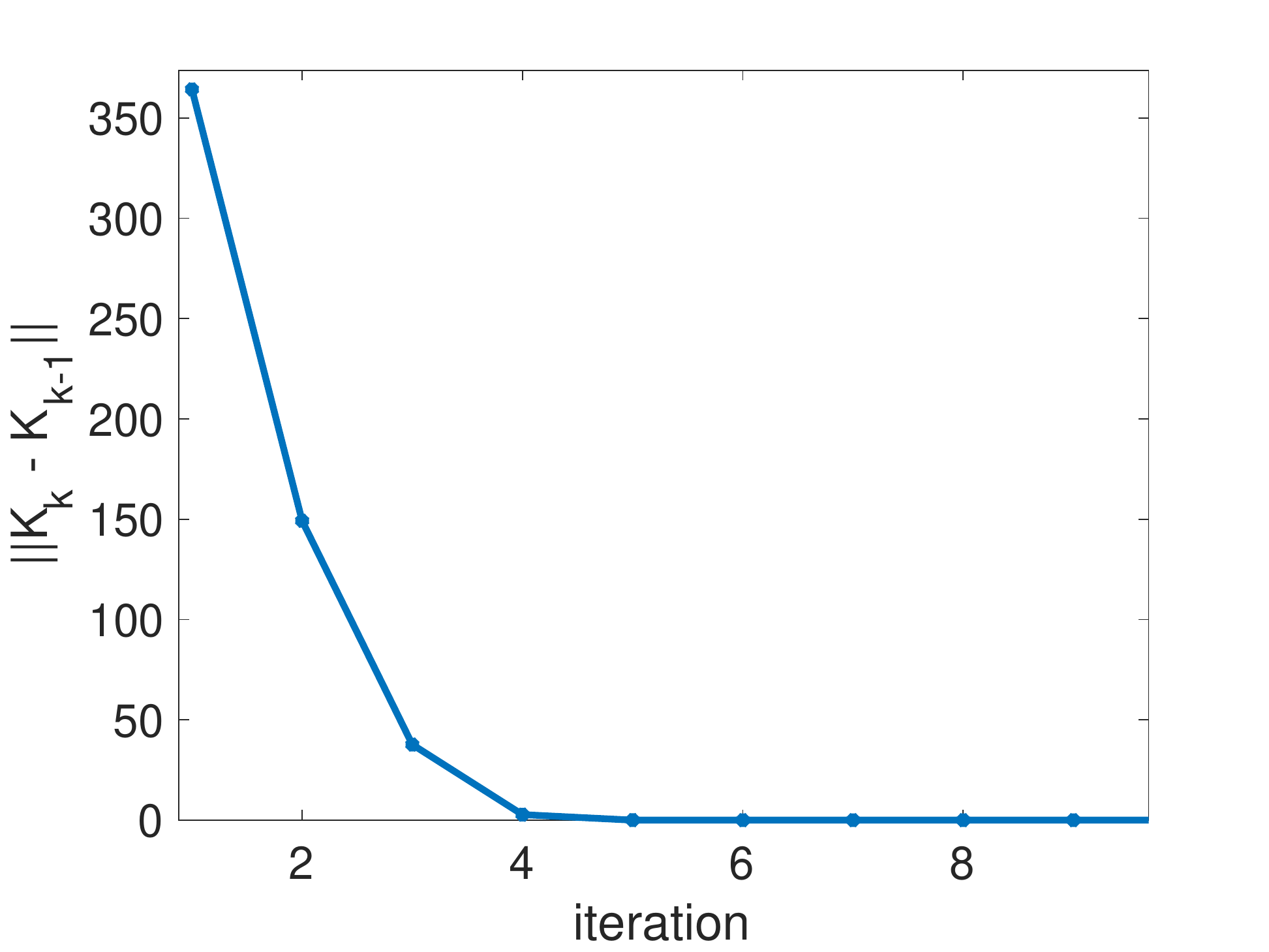} 
        \caption{\small{Convergence of $K$ using nominal RL (Algorithm 1)}}
        \label{fig:grid_covert_K}
    \end{minipage}
    \begin{minipage}{0.3\linewidth}
        \centering
        \includegraphics[width = \linewidth, height =2.7 cm]{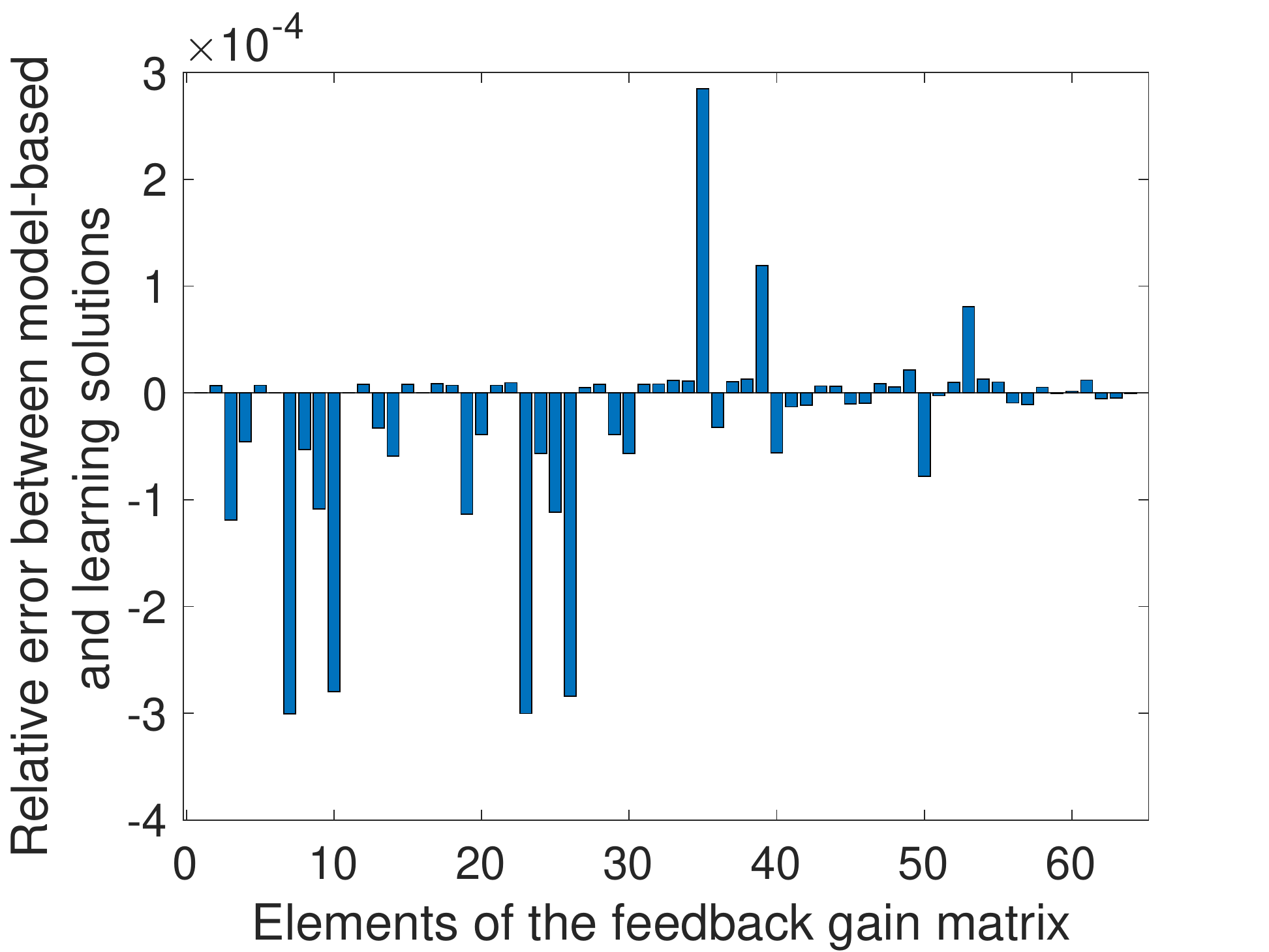} 
        \caption{\small{Error $\%$ of the RL based solution using Alg. 1 with respect to the model based solution}}
        \label{fig:grid_covert_error}
    \end{minipage}
    \begin{minipage}{0.3\linewidth}
        \centering
        \includegraphics[width = \linewidth, height =2.7 cm]{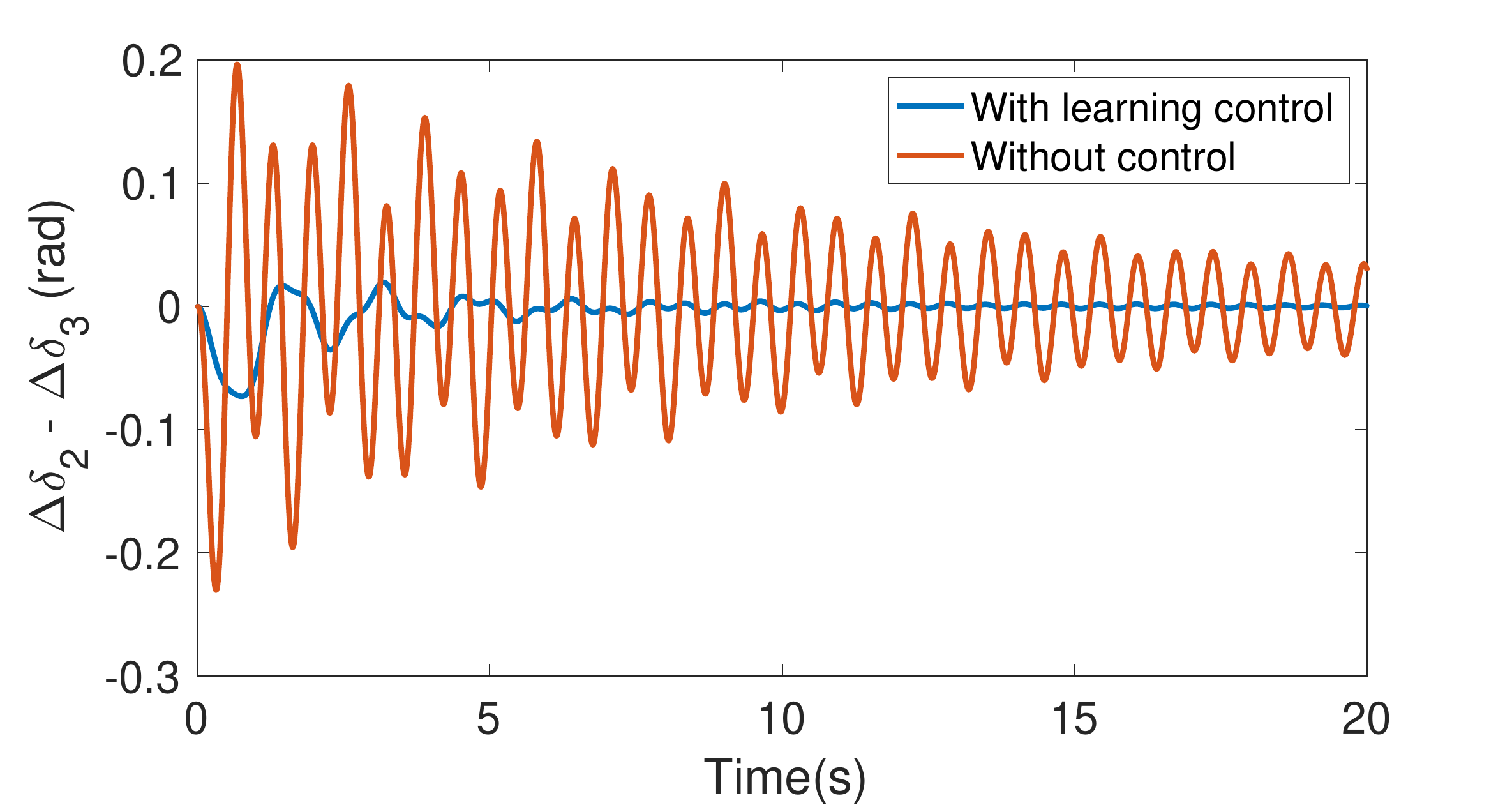} 
        \caption{\small{Improving the inter-area dynamic performance with the nominal RL control}}
        \label{fig:grid_covert_perf}
    \end{minipage}
    \begin{minipage}{0.3\linewidth}
        \centering
        \includegraphics[width = \linewidth, height =2.7 cm]{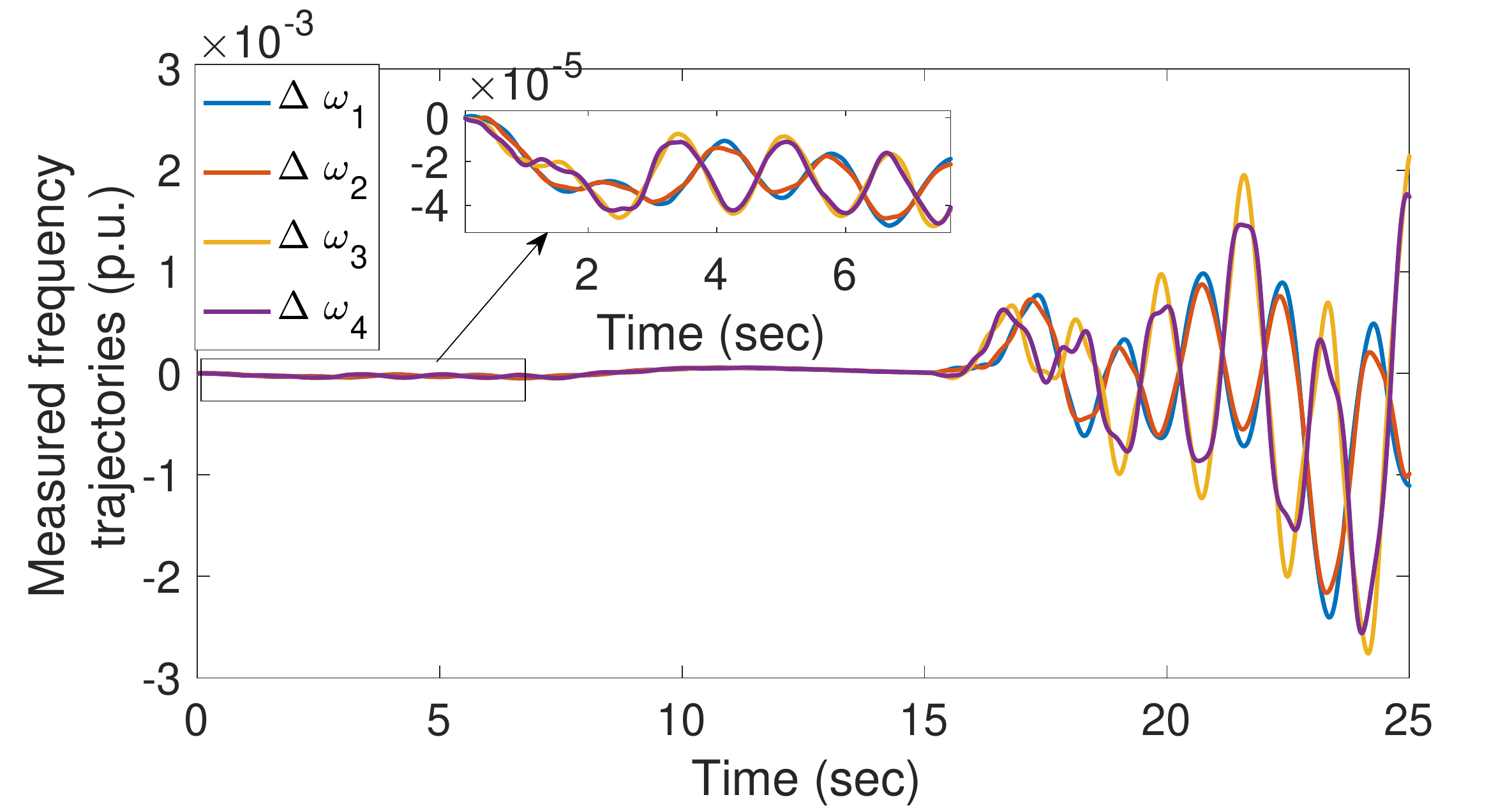} 
        \caption{\small{Measured frequencies during probing (till 7 s), control implementation, and attack at 15 s with ARRL}}
        \label{fig:grid_noncovert_meas}
    \end{minipage}
    \qquad
    \begin{minipage}{0.3\linewidth}
        \centering
        \includegraphics[width = \linewidth, height =2.7 cm]{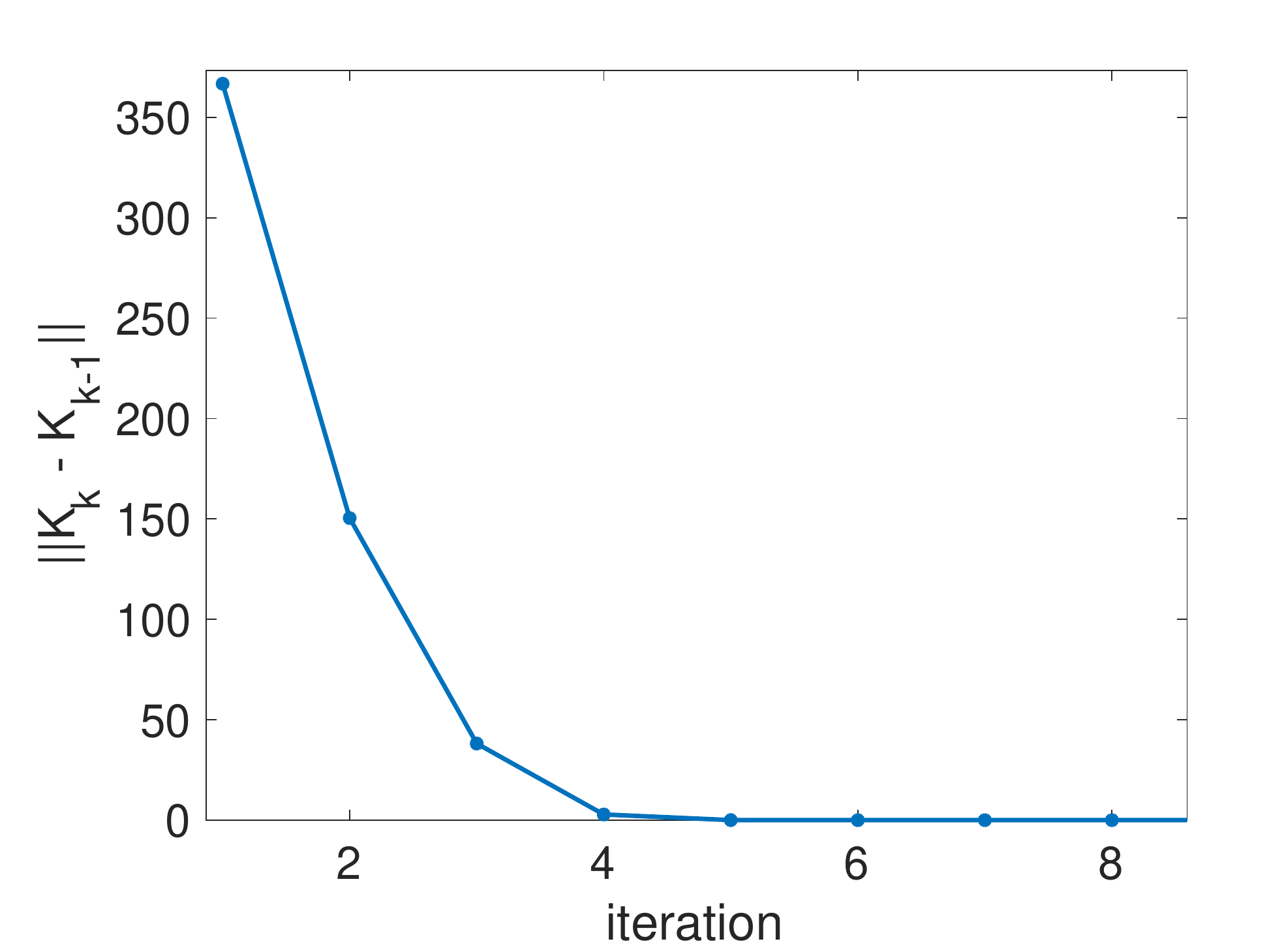} 
        \caption{\small{Convergence of the retrofitted controller following ARRL in Algorithm 2}}
        \label{fig:grid_noncovert_K}
    \end{minipage}
    \hspace{.4 cm}
    \begin{minipage}{0.3\linewidth}
        \includegraphics[width = \linewidth, height =2.7 cm]{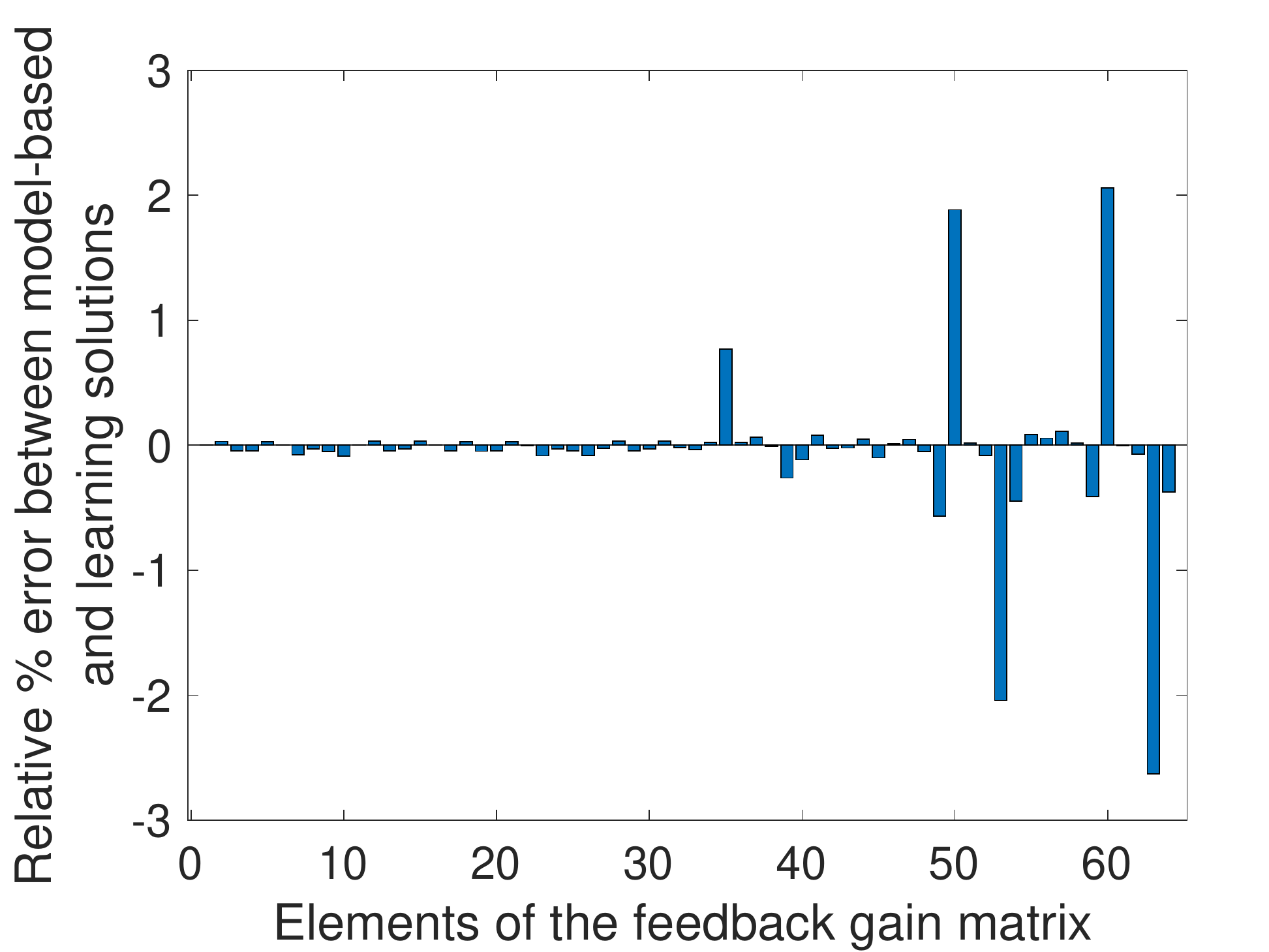} 
        \caption{\small{Error $\%$ of the RL based solution using ARRL (Alg. 2) with respect to the model based solution}}
        \label{fig:grid_noncovert_error}
    \end{minipage}
    \hspace{.55 cm}
    \begin{minipage}{0.3\linewidth}
        \centering
        \includegraphics[width = \linewidth, height =2.7 cm]{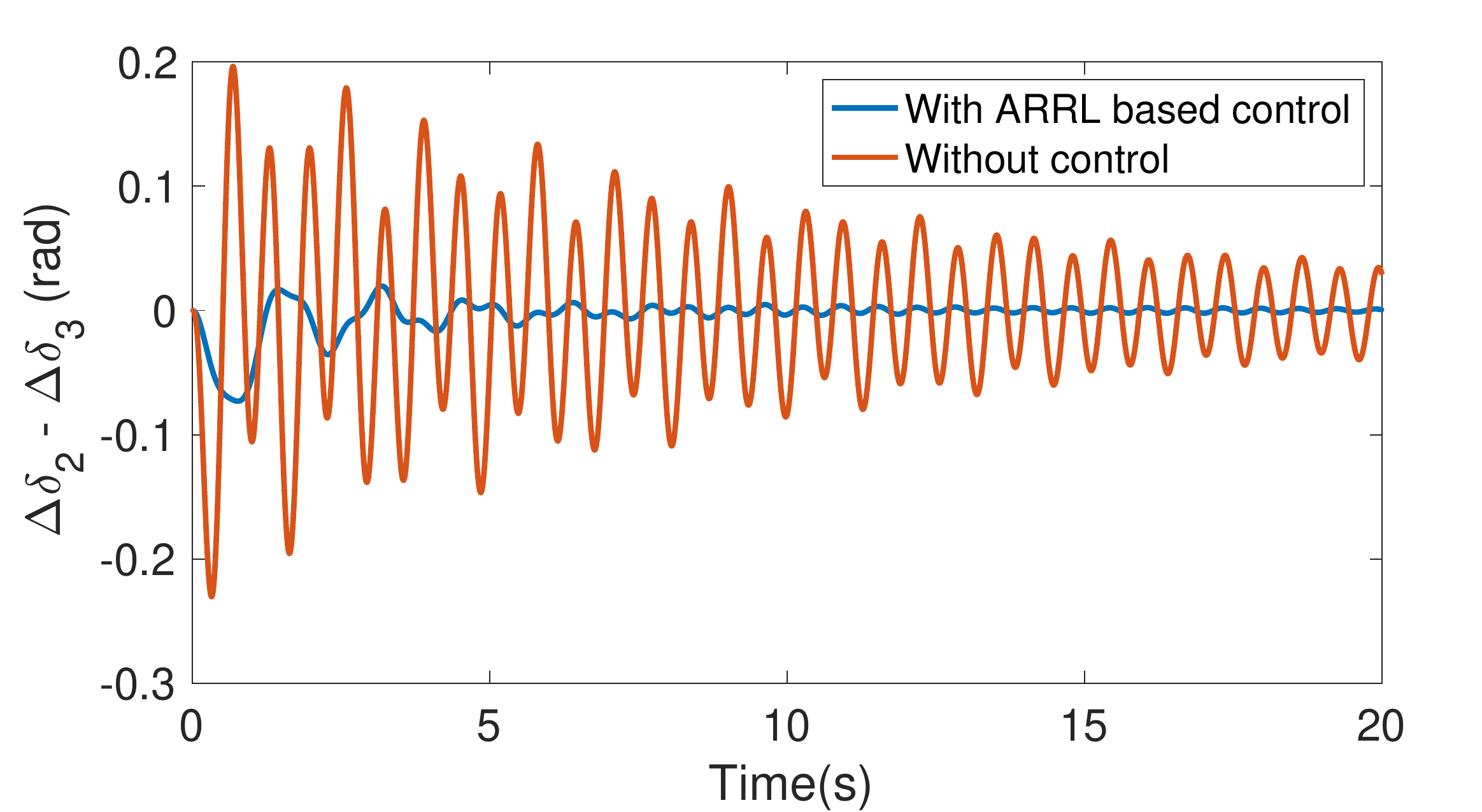} 
        \caption{\small{Improving the inter-area dynamic performance with the ARRL control}}
        \label{fig:grid_noncovert_perf}
    \end{minipage}
    \vspace{-.5 cm}
\end{figure*}
We consider a 13-bus Kundur power system model as shown in Fig. \ref{fig:Kundur}. To numerically simulate the model to generate data, we model generator dynamics by the \textit{flux decay} model. The state variables are denoted as $\delta(t)$, $\omega(t)$, $E(t)$, $E_{fd}(t)$ which are the generator internal angle in radians, speed deviation from nominal ($2\pi\times60$ radians/sec), internal voltage,  and excitation voltage in per unit, respectively. The supplementary control signal is added with the voltage reference of automatic voltage regulator (AVR) in the excitation dynamics. 
With a $100$ MVA base, the rated active power generation capacity of the generators G$1$-G$4$ are $7$ p.u., $7$ p.u., $7.16$ p.u., and $7$ p.u., respectively. The grid contains one inter-area mode ($0.61$ Hz) and two intra-area modes ($1.68,1.56$ Hz). 
The control design objective is to improve the inter-area and oscillatory dynamic performance of the grid. We have designed $Q$ such that it penalizes the relative difference between the generator angles (inter-area power flows depend on angular differences), and the energy associated with other generator states. In this model, we have $n=16, m=4$, and therefore, we explore for $5$ s with $0.01$ s time steps. The attack starts at $15$ s, and we simulate till $25$ s. Fig. \ref{fig:grid_covert_act} shows the actual frequency excursions during the learning and the attack which shows considerable perturbations caused by the adversary, however, in the measured frequencies, as in Fig. \ref{fig:grid_covert_meas}, the impact of the attack is not visible, making the attack covert. The fast convergence of the RL control update iterations are shown in Fig. \ref{fig:grid_covert_K}. The convergence is reached with high accuracy with respect to the model-based solutions as shown in Fig. \ref{fig:grid_covert_error}. The dynamic performance improvement of the nominal controller is shown in Fig. \ref{fig:grid_covert_perf}, where the angular difference between the generators $2$ and $3$ characterizes the oscillations in the tie-lines connecting buses $6$ and $12$. Subsequently, we test the attack resilient design with the malicious signal injection starting at $15$ s. The trajectory measurements are gathered for $8$ s, and thereafter, the control iterations of Algorithm 2 (ARRL) has been performed which results in fast convergence as shown in Fig. \ref{fig:grid_noncovert_K}. Most of the entries of $K$ matches closely with the model-based solutions, and only few of the entries are $2-3 \%$ off. However, when we test the dynamic performance of the learned ARRL control, we can see from Fig. \ref{fig:grid_noncovert_perf} that the optimal performance of the nominal RL design has been recovered. Moreover, in this scenario, the measured frequencies now capture the impact of the the malicious attack starting from $15$ s as in Fig. \ref{fig:grid_noncovert_meas}, therefore, the attacker cannot remain covert anymore.     
\section{Conclusions}
This paper discussed a secure learning control methodology which is resilient to adversarial actions from malicious agents. The attacker eavesdrops the learning process and estimate dynamic information of the CPS to conduct covert attack subsequently. In such scenarios, we have discussed a dynamic camouflaging technique during the learning to misguide the attacker without compromising the accuracy of learning of the optimal control. We have shown that by coupling the dynamic system with nonlinear static time-varying functions can provide one such dynamic camouflaging with adequate guarantees. Numerical simulations conducted on a consensus multi-agent system, and  on a power system model validates the algorithmic and theoretical considerations. 
\bibliographystyle{IEEEtran}
\bibliography{ref}

\begin{thebibliography}{10}
\providecommand{\url}[1]{#1}
\csname url@samestyle\endcsname
\providecommand{\newblock}{\relax}
\providecommand{\bibinfo}[2]{#2}
\providecommand{\BIBentrySTDinterwordspacing}{\spaceskip=0pt\relax}
\providecommand{\BIBentryALTinterwordstretchfactor}{4}
\providecommand{\BIBentryALTinterwordspacing}{\spaceskip=\fontdimen2\font plus
\BIBentryALTinterwordstretchfactor\fontdimen3\font minus
  \fontdimen4\font\relax}
\providecommand{\BIBforeignlanguage}[2]{{%
\expandafter\ifx\csname l@#1\endcsname\relax
\typeout{** WARNING: IEEEtran.bst: No hyphenation pattern has been}%
\typeout{** loaded for the language `#1'. Using the pattern for}%
\typeout{** the default language instead.}%
\else
\language=\csname l@#1\endcsname
\fi
#2}}
\providecommand{\BIBdecl}{\relax}
\BIBdecl

\bibitem{pasqualetti2015control}
F.~Pasqualetti, F.~Dorfler, and F.~Bullo, ``Control-theoretic methods for
  cyberphysical security: Geometric principles for optimal cross-layer
  resilient control systems,'' \emph{IEEE Control Systems Magazine}, vol.~35,
  no.~1, pp. 110--127, 2015.

\bibitem{dibaji2019systems}
S.~M. Dibaji, M.~Pirani, D.~B. Flamholz, A.~M. Annaswamy, K.~H. Johansson, and
  A.~Chakrabortty, ``A systems and control perspective of cps security,''
  \emph{Annual Reviews in Control}, vol.~47, pp. 394--411, 2019.

\bibitem{li2016event}
H.~Li, Z.~Chen, L.~Wu, H.-K. Lam, and H.~Du, ``Event-triggered fault detection
  of nonlinear networked systems,'' \emph{IEEE Transactions on Cybernetics},
  vol.~47, no.~4, pp. 1041--1052, 2016.

\bibitem{mousavinejad2018novel}
E.~Mousavinejad, F.~Yang, Q.-L. Han, and L.~Vlacic, ``A novel cyber attack
  detection method in networked control systems,'' \emph{IEEE transactions on
  cybernetics}, vol.~48, no.~11, pp. 3254--3264, 2018.

\bibitem{vamvoudakis2014detection}
K.~G. Vamvoudakis, J.~P. Hespanha, B.~Sinopoli, and Y.~Mo, ``Detection in
  adversarial environments,'' \emph{IEEE Transactions on Automatic Control},
  vol.~59, no.~12, pp. 3209--3223, 2014.

\bibitem{chong2019tutorial}
M.~S. Chong, H.~Sandberg, and A.~M. Teixeira, ``A tutorial introduction to
  security and privacy for cyber-physical systems,'' in \emph{2019 18th
  European Control Conference (ECC)}.\hskip 1em plus 0.5em minus 0.4em\relax
  IEEE, 2019, pp. 968--978.

\bibitem{yuan2013resilient}
Y.~Yuan, Q.~Zhu, F.~Sun, Q.~Wang, and T.~Ba{\c{s}}ar, ``Resilient control of
  cyber-physical systems against denial-of-service attacks,'' in \emph{2013 6th
  International Symposium on Resilient Control Systems (ISRCS)}.\hskip 1em plus
  0.5em minus 0.4em\relax IEEE, 2013, pp. 54--59.

\bibitem{bai2017data}
C.-Z. Bai, F.~Pasqualetti, and V.~Gupta, ``Data-injection attacks in stochastic
  control systems: Detectability and performance tradeoffs,''
  \emph{Automatica}, vol.~82, pp. 251--260, 2017.

\bibitem{zhu2013performance}
M.~Zhu and S.~Martinez, ``On the performance analysis of resilient networked
  control systems under replay attacks,'' \emph{IEEE Transactions on Automatic
  Control}, vol.~59, no.~3, pp. 804--808, 2013.

\bibitem{de2017covert}
A.~O. de~Sa, L.~F.~R. da~Costa~Carmo, and R.~C. Machado, ``Covert attacks in
  cyber-physical control systems,'' \emph{IEEE Transactions on Industrial
  Informatics}, vol.~13, no.~4, pp. 1641--1651, 2017.

\bibitem{barboni2020detection}
A.~Barboni, H.~Rezaee, F.~Boem, and T.~Parisini, ``Detection of covert
  cyber-attacks in interconnected systems: A distributed model-based
  approach,'' \emph{IEEE Transactions on Automatic Control}, 2020.

\bibitem{state_est}
Y.~{Mao}, S.~{Diggavi}, C.~{Fragouli}, and P.~{Tabuada}, ``Secure
  state-reconstruction over networks subject to attacks,'' \emph{IEEE Control
  Systems Letters}, vol.~5, no.~1, pp. 157--162, 2021.

\bibitem{satchidanandan2016dynamic}
B.~Satchidanandan and P.~R. Kumar, ``Dynamic watermarking: Active defense of
  networked cyber--physical systems,'' \emph{Proceedings of the IEEE}, vol.
  105, no.~2, pp. 219--240, 2016.

\bibitem{mo2015physical}
Y.~Mo, S.~Weerakkody, and B.~Sinopoli, ``Physical authentication of control
  systems: Designing watermarked control inputs to detect counterfeit sensor
  outputs,'' \emph{IEEE Control Systems Magazine}, vol.~35, no.~1, pp. 93--109,
  2015.

\bibitem{kanellopoulos2019moving}
A.~Kanellopoulos and K.~G. Vamvoudakis, ``A moving target defense control
  framework for cyber-physical systems,'' \emph{IEEE Transactions on Automatic
  Control}, vol.~65, no.~3, pp. 1029--1043, 2019.

\bibitem{schellenberger2017detection}
C.~Schellenberger and P.~Zhang, ``Detection of covert attacks on cyber-physical
  systems by extending the system dynamics with an auxiliary system,'' in
  \emph{2017 IEEE 56th Annual Conference on Decision and Control (CDC)}.\hskip
  1em plus 0.5em minus 0.4em\relax IEEE, 2017, pp. 1374--1379.

\bibitem{griffioen2020moving}
P.~Griffioen, S.~Weerakkody, and B.~Sinopoli, ``A moving target defense for
  securing cyber-physical systems,'' \emph{IEEE Transactions on Automatic
  Control}, 2020.

\bibitem{barto}
R.~Sutton and A.~Barto, \emph{Reinforcement learning - An introduction}.\hskip
  1em plus 0.5em minus 0.4em\relax MIT press, Cambridge, 1998, 1998.

\bibitem{vrabie1}
D.~Vrabie, O.~Pastravanu, M.~Abu-Khalaf, and F.~Lewis, ``Adaptive optimal
  control for continuous-time linear systems based on policy iteration,''
  \emph{Automatica}, vol.~45, pp. 477--484, 2009.

\bibitem{jiang1}
Y.~Jiang and Z.-P. Jiang, ``Computational adaptive optimal control for
  continuous-time linear systems with completely unknown dynamics,''
  \emph{Automatica}, vol.~48, pp. 2699--2704, 2012.

\bibitem{V17}
K.~Vamvoudakis, ``Q-learning for continuous-time linear systems: A model-free
  infinite horizon optimal control approach,'' \emph{Systems \& Control
  Letters}, vol. 100, pp. 14--20, 2017.

\bibitem{V18}
B.~Kiumarsi, K.~Vamvoudakis, H.~Modares, and F.~Lewis, ``Optimal and autonomous
  control using reinforcement learning: A survey,'' \emph{IEEE Trans. on Neural
  Networks and Learning Systems}, 2018.

\bibitem{MUKHERJEE_auto}
\BIBentryALTinterwordspacing
S.~Mukherjee, H.~Bai, and A.~Chakrabortty, ``Reduced-dimensional reinforcement
  learning control using singular perturbation approximations,''
  \emph{Automatica}, vol. 126, p. 109451, 2021. [Online]. Available:
  \url{http://www.sciencedirect.com/science/article/pii/S000510982030649X}
\BIBentrySTDinterwordspacing

\bibitem{de2019formulas}
C.~De~Persis and P.~Tesi, ``Formulas for data-driven control: Stabilization,
  optimality, and robustness,'' \emph{IEEE Transactions on Automatic Control},
  vol.~65, no.~3, pp. 909--924, 2019.

\bibitem{mukherjee2020reinforcement}
S.~Mukherjee and T.~L. Vu, ``Reinforcement learning of structured control for
  linear systems with unknown state matrix,'' \emph{arXiv preprint
  arXiv:2011.01128}, 2020.

\bibitem{fattahi2020efficient}
S.~Fattahi, N.~Matni, and S.~Sojoudi, ``Efficient learning of distributed
  linear-quadratic control policies,'' \emph{SIAM Journal on Control and
  Optimization}, vol.~58, no.~5, pp. 2927--2951, 2020.

\bibitem{dean2019sample}
S.~Dean, H.~Mania, N.~Matni, B.~Recht, and S.~Tu, ``On the sample complexity of
  the linear quadratic regulator,'' \emph{Foundations of Computational
  Mathematics}, pp. 1--47, 2019.

\bibitem{cps_vam}
Y.~{Zhou}, K.~G. {Vamvoudakis}, W.~M. {Haddad}, and Z.~P. {Jiang}, ``A secure
  control learning framework for cyber-physical systems under sensor and
  actuator attacks,'' \emph{IEEE Transactions on Cybernetics}, pp. 1--13, 2020.

\bibitem{rangi2020learning}
A.~Rangi, M.~J. Khojasteh, and M.~Franceschetti, ``Learning-based attacks in
  cyber-physical systems: Exploration, detection, and control cost
  trade-offs,'' \emph{arXiv preprint arXiv:2011.10718}, 2020.

\bibitem{kleinman}
D.~Kleinman, ``On an iterative technique for riccati equation computations,''
  \emph{IEEE Trans. on Automatic Control}, vol.~13, no.~1, pp. 114--115, 1968.

\bibitem{jiang_book}
Y.~Jiang and Z.-P. Jiang, \emph{Robust Adaptive Dynamic Programming}.\hskip 1em
  plus 0.5em minus 0.4em\relax Wiley-IEEE press, 2017.

\bibitem{mukherjee2020robust}
S.~Mukherjee, H.~Bai, and A.~Chakrabortty, ``On robust model-free
  reduced-dimensional reinforcement learning control for singularly perturbed
  systems,'' in \emph{2020 American Control Conference (ACC)}.\hskip 1em plus
  0.5em minus 0.4em\relax IEEE, 2020, pp. 3914--3919.

\bibitem{mukherjee2020imposing}
S.~Mukherjee and T.~L. Vu, ``Imposing robust structured control constraint on
  reinforcement learning of linear quadratic regulator,'' \emph{arXiv preprint
  arXiv:2011.07011}, 2020.

\end{thebibliography}
\end{document}